\documentclass{article}
\usepackage{fullpage}
\usepackage{amsfonts,amssymb}
\usepackage{amsthm}
\usepackage{tikz}
\usetikzlibrary{calc}
\usetikzlibrary{arrows}
\usepackage{tikz-3dplot}
\usepackage{color} 
\usepackage[fleqn]{amsmath}
\usepackage{amsthm}
\usepackage{amssymb}
\usepackage{amsfonts}
\usepackage{bbm}
\usepackage{epsfig}
\usepackage{times}
\usepackage{hyperref}
\usepackage{bm}
\usepackage{float} 
\usepackage{appendix} 
\usepackage{lscape}
\usepackage{cite}
\usepackage{mathrsfs}
\usepackage{appendix}
\usepackage{setspace}
\usepackage{mathtools}
\usepackage{mdframed}
\usepackage{authblk}
\theoremstyle{definition}

\begin{document}
	
	\title{Contextuality, Fine-Tuning and Teleological Explanation}	
	\author[1]{Emily Adlam} 
	
	\affil[1]{The University of Western Ontario}  
	
	\maketitle
	
	The study of contextuality has made great strides in recent years. 
On the mathematical side, we now have a rich understanding of the limits of quantum contextuality thanks to several powerful formalisms\cite{Cabello_2014, MUCKENHEIM1986337,Feintzeig_2017,Abramsky_2011}. On the conceptual side, there are ongoing discussions both about the reasoning that leads us to prefer non-contextual theories \cite{spekkens2019ontological, Acuna2021-ACUMHV} and the sorts of models which could naturally be expected to lead to contextuality \cite{rudolph2006ontological,harrigan2007ontological}. My aim in this article is to clarify how these various approaches to contextuality relate to one another and to explore what this combined body of work means for a scientific realist. Obviously I will not be able to mention every relevant result, but I aim to cover enough to get a clear picture of the conceptual implications of contextuality. 

I begin by assessing various proposals for the source of the intuition that there is something problematic about contextuality, ultimately concluding that contextuality is best thought of in terms of fine-tuning. I then argue that as with other fine-tuning problems in quantum mechanics, this behaviour can be understood as a manifestation of teleological features of physics; I therefore suggest that contextuality can be regarded as evidence for the `all-at-once' approach advocated in refs \cite{QMG, adlam2020tsirelsons, Wharton}. Finally I discuss several formal mathematical frameworks that have been used to analyse contextuality and consider how their specific results should be interpreted.

In the course of this discussion I prove several new mathematical results. I use the methods of ref \cite{Cavalcanti_2018} to demonstrate that preparation contextuality is a form of fine-tuning; I show that measurement contextuality can be explained by appeal to a global constraint forbidding closed causal loops; and I demonstrate how negative probabilities arise from a classical ontological model together with an epistemic restriction.

\section{A Note on Terminology \label{intro}}

Throughout this article we will be dealing with \emph{contextuality scenarios}. A contextuality scenario is a set of measurements together with a specification of contexts, where a context is a set of measurements which can be performed together, i.e. they are all `compatible.' When we are dealing with quantum mechanics we say that measurements are compatible if the corresponding measurement operators commute, but what if we want to understand contextuality in a more general operational context without presupposing the mathematical formalism of quantum mechanics? In this article, we will take it that  a set of measurements $\mathbb{C}$ on a (possibly composite) system $S$ are compatible if for any possible preparation of $S$, if we perform all the measurements in $\mathbb{C}$ on $S$, the probability distributions over the outcomes of each individual measurement are independent of the order (defined in any convenient reference frame) in which the measurements are performed. Clearly a set of commuting quantum measurements will always have this property. We will say that a context is \emph{maximal} if it is not possible to add any other measurement (other than the trivial one)  to the set whilst retaining this property. Note that in the classical world all measurements are (in principle) non-disturbing, so for a classical system all possible measurements form a single maximal context.

There are two main ways in which we may form `contexts' in quantum mechanics and other non-classical theories. If we are dealing with two or more distinct physical systems, we can perform measurements on each system individually, and in quantum mechanics the no-signalling theorem ensures that performing measurements on one system does not change the probability distributions for measurements on the other systems, so the probability distributions over outcomes are independent of the order of measurement in any reference frame and thus each possible combination of individual measurements defines a context. Clearly  we can keep the measurement on one system  constant while changing the choice of measurement on other systems, so a single measurement can appear in several different contexts, and the no-signalling theorem ensures that in quantum mechanics the probability distribution over the outcomes of a measurement is independent of the context.

In the case where we are dealing with only one physical system, a context usually refers to a multiple-outcome measurement such as $  P = \{  | 0 \rangle \langle 0 |, | 1 \rangle \langle 1 |, | 2 \rangle \langle 2 |\}$. Although we may be accustomed to thinking of $P$ as a single measurement in its own right, it can equivalently be regarded as a set of three measurements $\{ | 0 \rangle \langle 0 |, \:  \: \mathbb{I} - | 0 \rangle \langle 0 |\}$, $\{ | 1 \rangle \langle 1 |,  \: \mathbb{I} - | 1 \rangle \langle 1 |\}$ and $\{ | 2 \rangle \langle 2 |,  \: \mathbb{I} - | 2 \rangle \langle 2 |\}$. The probability to obtain the result $| 0 \rangle \langle 0 |$ to the measurement $\{ | 0 \rangle \langle 0 |,  \: \mathbb{I} - | 0 \rangle \langle 0 |\}$, is the same regardless of the order in which we perform these three measurements, and likewise for the other two measurements, so $P$ defines a (maximal) context. For the sake of clarity, in this article we will use `measurement' to refer to a two-outcome test of the form $\{ | 0 \rangle \langle 0 |,  \: \mathbb{I} - | 0 \rangle \langle 0 |\}$ and we will use `context' to refer to more complex operations like $P$ which can be regarded as sets of compatible measurements.

 In general,  for quantum systems of dimension greater than two a  measurement may appear in more than one context: for example the measurement $\{ | 0 \rangle \langle 0 |,  \: \mathbb{I} - | 0 \rangle \langle 0 |\}$ appears in the context $ \{  | 0 \rangle \langle 0 |, | 1 \rangle \langle 1 |, | 2 \rangle \langle 2 |\}$  and also the context $ \{  | 0 \rangle \langle 0 |,  \frac{1}{2}(| 1 \rangle + | 2 \rangle )( \langle 1 | + \langle 2 | ),  \frac{1}{2}(| 1 \rangle - | 2 \rangle )( \langle 1 | - \langle 2 | )\}$. Quantum mechanics has the feature that the probability distribution over the outcomes of a quantum measurement does not depend on which other compatible measurements are being performed simultaneously, i.e. once again the probability distribution over the outcomes of a measurement is independent of the context; I will refer to this property as `Gleason's property,' because it plays an important role in Gleason's theorem\cite{10.2307/24900629}. Many theories that we are familiar with satisfy Gleason's property trivially - e.g. because no measurement appears in more than one context, as in two-dimensional quantum mechanics, or because there is only one maximal context, as in classical mechanics. But quantum mechanics for systems of dimension greater than two satisfies Gleason's property in a non-trivial way, since there are multiple maximal contexts and some measurements appear in more than one maximal context, so it would be logically possible for a given measurement to have different probabilities in different maximal contexts.

\section{What's the problem with contextuality? \label{problem}}

 In its first incarnation, contextuality was associated with the idea that quantum measurement outcomes correspond to definite properties of systems - so if a system has the property associated with an outcome of $A$ to the measurement $\{ A,  \: \mathbb{I} -  A\}$, then whenever we perform this measurement we are certain to get the outcome $A$, regardless of what other measurements we are performing at the same time. A hidden variable model in which measurement works this way can be said to obey \emph{deterministic non-contextuality}. It is trivial to find an assignation of definite properties satisfying deterministic non-contextuality in a setting where each measurement   can occur in only one context, but   in quantum mechanical systems of more than two dimensions a given measurement   can occur in several different possible contexts, and the Kochen-Specker theorem proves that for certain collections of measurements for quantum systems of more than two dimensions, there is no possible way of picking a set of outcomes which correspond to `definite properties'\cite{KochenSpecker} - if we insist on assigning deterministic outcomes to measurements, we will sometimes have to assign them in such a way that we are certain to get  an outcome of $A$ to the measurement $\{ A,  \: \mathbb{I} -  A\}$ when we perform it in some context   and we are certain   \emph{not} to get  an outcome of $A$ to the measurement $\{ A,  \: \mathbb{I} -  A\}$ if we perform it in some other context.  So quantum mechanics does not obey deterministic non-contextuality and therefore it is not possible to think of quantum measurement outcomes as simply describing pre-existing properties of systems.

For realists, there is an obvious response to be made here: the demand for deterministic non-contextuality is  a confusion arising from the naive classical view that measurements  reveal pre-existing properties, and its lesson is simply that we should avoid taking the mathematical structure of the theory so literally. Indeed, an earlier result by von Neumann had already demonstrated the impossibility of deterministic hidden variable models whose variables are elements of a Hilbert space\cite{von2018mathematical}, and Ancuna has recently argued that Gleason's theorem leads to a similar conclusion\cite{Acuna2021-ACUMHV}. As discussed at length in the subsequent literature\cite{Bub_2010, Jammer1974-JAMTPO-10, dieks2018von,Acuna2021-ACUMHV} these various results provide abundant evidence against straightforward deterministic hidden variable models which read ontic structure directly off mathematical structure,  but they don't rule out the possibility of more general hidden variable models - indeed, since the Kochen-Specker theorem was proved a number of models, such as Spekkens' toy model\cite{Spekkensepistemic}, have been proposed to demonstrate that many of the predictions of quantum mechanics can be derived from `psi-epistemic' models where the underlying ontic structure is different from the mathematical structure. Therefore these results are no great threat to the broader realist project or indeed to the possibility of determinism.  

However, there exist more sophisticated approaches to contextuality. Spekkens has popularised a form of generalized contextuality based on the idea that  operationally equivalent scenarios should correspond to the same underlying ontic reality\cite{Spekkenscontextuality}; I will henceforth refer to this idea as the Ontic Equivalence Principle (OEP). To make OEP precise, Spekkens employs the ontological models framework (see appendix \ref{ontological}), in which each possible preparation is associated with an epistemic state distribution $\mu_P$ over ontic states and each measurement $M$ and outcome  $k$ is associated with a response function $\xi^{k, M}$.  Spekkens uses this framework to identify several different sorts of contextuality:  for now let us focus on measurement contextuality, which can be regarded as the non-deterministic generalisation of Kochen-Specker contextuality. An ontological model is said to be measurement non-contextual if a quantum mechanical measurement operator is represented in the ontological model by the same response function in every possible context, and Spekkens has shown that under plausible assumptions, no ontological model for quantum mechanics can be measurement non-contextual\cite{Spekkenscontextuality}.

Because OEP is formulated in operational terms, Spekkens' approach to contextuality avoids the pitfall of naively reading structure off a particular mathematical formulation. But what exactly is the status of OEP, and why should we find it convincing? A number of different arguments have been proposed. First, it  has been suggested\cite{Cavalcanti_2018, spekkens2019ontological, schmid2020unscrambling} that OEP is a version of Leibniz's Principle of the Identity of Indiscernibles (IOI). However, IOI is usually understood as a metaphysical thesis about numerical identity: Leibniz claimed it was a conceptual, a priori truth that there cannot be two numerically distinct physical objects, or two numerically distinct possible worlds, which share all the same properties\cite{sep-identity-indiscernible}. Whereas OEP does not seem to be intended as a conceptual truth, as no one  is arguing that  it is \emph{impossible} for  operationally equivalent phenomena to differ at the ontic level. Indeed, in order to make such an argument one would probably have to adopt a fairly strong form of idealism which entails that reality supervenes on our observations, in which case there would be little sense in postulating a set of underlying ontic states at all. In fact, the proponents of OEP usually present it as a \emph{methodological} principle about how to construct models: it is supposed to be a useful but ultimately fallible guide to action, rather than an exceptionless conceptual truth. 

Moreoever, OEP also does not seem to be intended as a thesis about numerical identity. Say for example that some measurement  $E$ is performed in two different contexts. Regardless of whether or not $E$ can be represented by the same response function in the two contexts, there is no question that these two measurements are numerically distinct  - they take place in different spacetime locations, they involve different measurement procedures, and they are performed on different physical systems. Thus  OEP and IOI have distinctly different subject matter: IOI deals with the metaphysics of identity while OEP deals with the \emph{epistemology} of whether we could   ever have good reason to suppose that operationally equivalent processes are ontologically distinct.

This is not just  nit-picking, because it follows from this observation that the types of argument in which IOI is usually employed are not available in the contextuality case. For example, Spekkens compares the discussion over contextuality to Leibniz's famous argument against the existence of absolute space, where he asks us to consider  two possible worlds such that the whole of the universe is shifted by some distance in the first universe relative to its position in the second, and then invokes IOI to argue that these are not really different possible worlds.\footnote{I take no position here on whether Leibniz's argument is right; I only mean to draw attention to its particular form} But in the  contextuality case this sort of  argument can't be made, for if there are two possible worlds such that in one world an agent performs a measurement in one context and in the other world the agent performs the same measurement in a different context, there can be no doubt that these are distinct worlds, and therefore IOI is not relevant to the question of whether or not the response functions associated with the measurement in these two contexts are the same or not. Moreover the fact that IOI is supposed to be known a priori means that (if its status as a priori knowledge is accepted) it needs no further justification; but OEP is certainly not an a priori conceptual truth and therefore some other justification must be offered for it,  so comparisons to IOI don't really do anything to explain why we should find OEP plausible. 

It has also been suggested\cite{spekkens2019ontological} that OEP can be regarded as a generalization of Einstein's ideas - for example, his belief that we should try to rid our theories of `asymmetries which do not appear to be inherent in the phenomena,'\cite{Einstein} which was an important motivation for the theory of Special Relativity. Einstein's argument here is a  relational one: he is pointing out a mismatch between the phenomena and the model which can be removed simply changing the model. We can therefore understand him as invoking a version of Ockham's razor - since this structure can be removed by a simple change of model, it is superfluous and therefore we shouldn't attach ontological significance to it. But the same argument can't be made in the case of quantum contextuality, as we have a number of proofs to the effect that no model of quantum mechanics can be non-contextual, and so    it appears that  this particular asymmetry \emph{is} in fact inherent in the phenomena: since there can be no non-contextual model of quantum mechanics, contextuality must be playing some non-trivial role, and therefore Ockham's razor has every reason to spare it. Yet the various proofs that contextuality is an essential feature of quantum mechanics have not put an end to the discussion, and there still seems to be a sense that it would be desirable to get rid of contextuality if that were possible. This sense can't be based on an appeal to Ockham's razor, since it is provably not the case that contextuality is `superfluous structure' - so what is the real motivation here? 

One way of answering this question would be to argue that proofs of quantum contextuality depend on a set of assumptions about the nature of the underlying model, and therefore quantum contextuality is likewise relativized - not in this case to a single model, but to a \emph{class} of models. In particular, Spekkens' proof of measurement contextuality depends on representing measurements as two-way interactions between a state and an object representing the individual measurement, i.e. a response function, so perhaps we can get rid of the contextuality by moving to models which don't represent measurements in this way. For example, in Rudolph's `marble-world' model, the individual measurements  and the ontic states are both represented by projectors in the state space, and the result of a measurement is simply the projector which is closest to to the current ontic state, so when we perform $n$ compatible measurements, the outcome is determined by an $(n + 1)$-way interaction between the ontic state vector and all of the $n$ projectors. Although it is not possible to represent measurements by a unique response function in marble-world,  it is nonetheless the case that in the model they are represented by a unique \emph{object}, i.e. a vector in the state space, so the structure of the model mirrors the relevant operational equivalences and therefore marble-world is in a general sense non-contextual relative to its own conception of the content of reality. The box below gives some more details on marble-world, and several other models in which contextuality seems to arise in a natural way. In the context of these sorts of models it seems very reasonable that the result of a measurement should depend on which compatible measurements are also being performed - so reasonable, in fact, that one might begin to wonder why anyone would ever have thought otherwise.

But in fact, there \emph{is} a good reason that to think otherwise: Gleason's property. Since Gleason's property is precisely the `operational equivalence' which grounds the intuition that measurement outcomes should be represented identically in every context in the first place, any putative explanation for contextuality must also satisfactorily explain Gleason's property. And marble-world does not automatically satisfy the Gleason property: in order to do so, the model would have to guarantee that for any pair of contexts of the form  $M_1 = \{ A, B, C\}$ and $M_2 = \{ C, D, E\}$, and for any probability distribution over ontic states produced by a valid preparation, the probability that the state lies in the region closer to $C$ than to $A$ or $B$ is the same as the probability that it lies in the region closer to $C$ than to $D$ or $E$. No simple choice for probability distributions over ontic states is likely to have this property, and thus although marble-world may offer a natural way to explain the dependence of the outcome on the context, it does \emph{not} explain why the dependence of the outcome on the context should be hidden by the Gleason property.  

There are possible ways to rectify this - for example, by tinkering with the probability distributions to ensure that Gleason's property is obeyed after all. But it seems likely that the parameters of the model would have to be very carefully chosen to achieve this, which is to say, the model would have to be fine-tuned. For example, the de Broglie-Bohm theory manages to reproduce the empirical results of quantum theory only because the de Broglie-Bohm particles are always assumed to start out in a very specific distribution, and it has frequently been argued that this special fine-tuned distribution is in need of explanation\cite{Valentini_2005, valentini2019foundations, Durr_1992}. And in fact, this is a generic feature. The obvious way to explain  Gleason's property is to say that measurements are associated with stable underlying properties such that the probability of obtaining the outcome $E$ to the measurement $\{ E,  \: \mathbb{I} -  E \}$ after preparation $P$ is equal to the relative frequency of the  property corresponding to $E$ in a large ensemble of systems prepared according to $P$, which is naturally independent of  context. But quantum contextuality blocks this sort of account, so instead we are forced to postulate models such that for any given ontic state the probability  distribution over the outcomes of a given measurement may depend on the  context, but when we average over ontic states according to the distributions associated with possible preparations, the probabilities  will no longer depend on  contexts. Yet any model which succeeds in `hiding' the dependence on context in this conspiratorial way will necessarily look very fine-tuned. 

 Cavalcanti has given a formal proof of the link between contextuality and fine-tuning by applying the framework of causal models to contextuality scenarios\cite{Cavalcanti_2018}. This involves attempting to describe causal influences between the variables involved in the scenarios and/or some set of `latent' variables in terms of a causal model, and saying that a causal model is faithful (i.e. not fine-tuned) if its causal graph exhibits a d-separation between any two variables which are conditionally independent, where (roughly speaking) two variables are d-separated if there is no direct causal path between them. Cavalcanti applies the framework to no-disturbance phenomenon - i.e. pairs of measurements where the choice of measurement setting for one measurement has no effect on the probability distribution over the outcomes for the other measurement and vice versa. Clearly tests of two-party Bell inequalities are no-disturbance phenomena, but standard measurement contextuality scenarios can also be put in this form - for example, given two maximal contexts   $\{ A, B, C\}$ and $\{ A, D, E\}$, we can let the first measurement be $\{ A,  \: \mathbb{I} -  A\}$ and then subsequently we choose either the two measurements $\{ B,  \: \mathbb{I} - B\}$, $\{ C,  \: \mathbb{I} - C\}$  or the two measurements $\{ D,  \: \mathbb{I} - D\}$, $\{E,  \: \mathbb{I} - E\}$. Due to Gleason's property, the result of  the first measurement doesn't depend on the choice we make in the second measurement, so this is indeed a no-disturbance phenomenon. Cavalcanti demonstrates that if a no-disturbance experiment is constructed from a  collection of measurements which together violate a Kochen-Specker inequality, then when we try to represent their behaviour with a causal model, it is necessarily the case that the model is not faithful. This confirms the conclusion we reached in the above discussion: the existence of Kochen-Specker contextuality does indeed entail that any causal model for the phenomena in question must be fine-tuned.  
 
 It's important to note that Cavalcanti's fine-tuning result does not make any assumptions about the nature of the latent variables or the form of the causal relations postulated by the model - the fine-tuning proof requires only the assumption that there exists \emph{some} causal account of the correlations, and therefore it is not relativized to any particular theoretical framework. Of course we have the option to move to a \emph{non-causal} model, and indeed this is exactly what I will advocate, but nonetheless it seems clear that the existence of contextuality is telling us something profound about reality itself, not merely about our choice of representation.

So in fact, I suggest that the appropriate way to motivate OEP is not by appeal to IOI, Ockham's razor, or superfluous structure, but in terms of our preference for models which avoid fine-tuning. The reason operationally equivalent situations should correspond to the same underlying ontic reality is simply that in cases where we have two or more ontic realities associated with the same operational statistics, the model will generally have to be fine-tuned in order to precisely cancel out dependencies so that no differences between these ontic realities are detectable.  Physicists tend to take the presence of fine-tuning as an indication that something further must be added to the model to explain the fine-tuning\cite{Carroll2,2015arXiv151003706A,sep-fine-tuning}, so we have good reason to pay special attention to places in physics where OEP seems to fail. Clearly this is a heuristic principle rather than an exceptionless rule: there might be special sorts of models where the multiplicity of the underlying ontic realities does not in fact look like fine-tuning, and in any case fine-tuned models are not logically impossible. But nonetheless, when we come across fine-tuning in physics it is considered good practice to at least make some attempt to explain it, so the existence of contextuality does indicate that there is some work to be done.

\begin{mdframed}
	
	\textbf{Measurement Device Configuration}
	
	\vspace{2mm}

In ref \cite{harrigan2007ontological} Rudolph and Harrigan argue that contextuality can be attributed to the `completely natural arrangement that the interactions of (the measurement device) and (the system) depend on the configuration of (the measurement device),' where the configuration of the measurement device is understood to include information about which other compatible measurements are being simultaneously performed. And indeed, this suggestion seems very natural - after all, we already know that this interaction must depend in some way on the configuration of the measurement device, since it is the configuration of the measurement device which determines the possible outcomes available, so it is not much of a leap to suppose that the response function may  depend in other ways on the choice of setting. 

However, this approach seems to work best in the case where the full set of compatible measurements are combined into a single measurement, e.g. $ \{  | 0 \rangle \langle 0 |, | 1 \rangle \langle 1 |, | 2 \rangle \langle 2 |\}$. It's less clear how it would apply if we first performed $ \{  | 0 \rangle \langle 0 |,  \: \mathbb{I} -  | 0 \rangle \langle 0 |\}$ and then subsequently decided which compatible measurements to perform - is our later choice supposed to have a retrocausal influence on the setting of the measurement device for the first measurement? Alternatively, one could suppose that there is a fixed, context-independent response function for the first measurement and then either the ontic state or the response functions for the subsequent measurements depend on the measurements which have already been made; so the `measurement setting' approach is only supposed to apply when the measurements are performed simultaneously. 

	\vspace{2mm}
\textbf{De Broglie-Bohm}
	\vspace{2mm}
	
The de Broglie-Bohm interpretation\cite{holland1995quantum} may be regarded as an example of the kind of model that Rudolph and Harrigan have in mind. In this picture, quantum systems are associated with de Broglie-Bohm particles having definite positions which are guided through space by the wavefunction. During a measurement, the  system being measured becomes entangled with some variables of the measuring device, so that the velocity of the de Broglie-Bohm particles of the system come to depend on the positions of the de Broglie-Bohm particles of the measuring device, and thus the result of the measurement may depend on the full configuration of the measurement device, including which other measurements are being performed simultaneously. Indeed, it is shown in ref \cite{tastevin2021outcomes} that in many cases the configuration of the measuring device is the main influence on the outcome, as we might expect given that the measuring device is usually much larger than the measured system and thus it dominates the interaction. 

The de Broglie-Bohm model can also deal with the  case where the measurements composing a context are performed sequentially rather than simultaneously:  the ontic state (i.e. the spatial distribution of the de Broglie-Bohm particles) is altered by the first measurement, and thus the later measurements naturally have a different ontic representation.

\vspace{2mm} 
	\textbf{Marble-world}

\vspace{2mm}

Another natural way to explain the dependence of measurement outcomes on context is to imagine that rather than being collections of properties, quantum states are like \emph{preference orderings}: that is to say, a state corresponds to an ordered list of all possible measurement outcomes such that, for any possible measurement, the highest-ranked outcome is the one that will occur. So the measurement $\{ A, B, C\}$ is to be understood not as asking `which of these three properties do you currently have?' but rather `which of these options do you prefer?' - and of course, when you are asked to choose a favourite amongst a set of options,  whether or not you choose option $A$ will usually depend on what other options are available. 

Rudolph suggests an explicit model of this kind, which I   refer to as `marble-world.'\cite{rudolph2006ontological} In this model, a quantum state corresponds to a probability distribution over ontic states, which are represented as projectors in a complex projective space. Measurements are also represented by projectors, and when we perform a set of compatible measurements, the outcome which occurs is simply the one associated with the measurement projector which happens to be closest to the current ontic state: Rudolph suggests thinking  of the ontic state as a marble on a sphere, which is attracted towards all of the measurement vectors and thus ultimately moves to the closest one. 

Rudolph's model does not perfectly reproduce the quantum mechanical statistics, but it is nonetheless suggestive. In particular, it is straightforward to see how contextuality can arise under these circumstances; given two measurements $M_1 = \{ A, B, C\}$ and $M_2 = \{ C, D, E\}$, it could easily be the case that $\lambda$ is closer to $C$ than to $A$ or $B$, but closer to $D$ than to $C$, so the result of $M_1$ will be $C$ but the result of $M_2$ will not be $C$. This model is clearly KS-contextual, and it is also measurement contextual in Spekkens' sense - if we write it down as an ontological model then the response function associated with $C$ must satisfy $\xi^{C, M_1}( \lambda  )= 1$ but $\xi^{C, M_2}( \lambda  )= 0$. But the contextuality arises in a very natural way, because measurements in marble world amount to asking systems to  `pick the best option.'

Again, it is not entirely clear how the  marble-world model is meant to work in the case where the measurements constituting a context are performed sequentially: if the measurements are temporally separated then the complex projective space in which the relevant vectors live can't actually be located at the spacetime locations of the measurements, since then we would never have the full set of vectors present at once, so if the attraction model is to be taken literally, marble-world would seem to entail the existence of some sort of platonic extra-temporal space in which these interactions can take place. Alternatively, as before we can suppose that   either the ontic state or the response functions for the subsequent measurements depend on the measurements which have already been made, so the vector attraction account applies only when all the measurements are being performed simultaneously.

\end{mdframed}

\subsection{Preparation Contextuality \label{prepfine}}

I have argued that violations of OEP are  best understood as  fine-tuning problems. To test this intuition, let us now see if it is true for another form of contextuality identified in Spekkens' analysis. Again using the ontological models framework, Spekkens specified that an ontological model is preparation non-contextual if for every set of preparation procedures which all produce the same quantum state, all of the preparations in the set are represented by the same probability distribution over ontic states in the ontological model; otherwise it is preparation contextual\cite{Spekkens}. Moreoever,  Spekkens proved that any ontological model which reproduces the observable results of quantum mechanics must be preparation contextual, subject to reasonable assumptions about the way ontic states behave under composition.

To see that preparation contextuality is indeed a fine-tuning problem in the causal modelling sense, observe that a scenario exhibiting preparation contextuality can be thought of as a no-disturbance phenomenon: given a set of preparations which all prepare the same quantum state, we select and perform one preparation and then measure the resulting state, and of course the result of this measurement will be independent of the choice of preparation. Thus we can apply Cavalcanti's methods to show that any faithful causal model for such a scenario is factorisable. But it follows from the definition of preparation contextuality that a preparation contextual ontological model for such a scenario will not be factorisable; so if a preparation contextual ontological model is turned into a causal model by treating the ontic variables as latent variables, the resulting causal model can't  be faithful. Thus we have shown that any preparation contextual ontological model is necessarily fine-tuned, so we are now able to see what is `wrong' with preparation contextual models: they are suspect because they have to be fine-tuned in order to hide the dependence of measurement outcomes on the choice of preparation. The details of this proof can be found in appendix \ref{appprepfine}.

It might have been tempting to think that we could get around Spekkens' preparation contextuality result by observing that the proof is based on the assumption that measurement results can depend only on facts about the present ontic state, whereas in a more general setting measurement results might depend on facts about the past and future as well. However, Costa and Shrapnel have used the process matrix formalism to demonstrate that even if we allow non-standard temporal orderings, any model which reproduces the observable results of quantum mechanics must exhibit `process contextuality,' which can be understood as a generalisation of preparation contextuality for scenarios with  non-standard causal ordering: a process `captures those physical features responsible for generating the joint statistics for a set of events, independently of the choice of local instruments.'\cite{Shrapnel_2018} And in fact this is exactly what we would expect if indeed preparation contextuality is a fine-tuning problem in the causal modelling sense, because causal models make no assumption about the temporal relationships between the variables in the model; in particular, there is no assumption that the latent variables are located to the past of the measurements or that their effect is spatially or temporally local, and therefore it is to be expected that simply relaxing some assumptions about the spacetime location of the latent variables will not get rid of the fine-tuning problem or the contextuality associated with it. 

\subsubsection{Biased and unbiased counterfactual outcomes}

To get a clearer understanding of the nature of the fine-tuning involved in preparation contextuality, it's helpful to consider the case of an ontological model where all the response functions are deterministic. In such a model, the ontic state $\lambda$ must record, for each possible context $C$, which set of outcomes $O$ will occurs if we perform the set of measurements in $C$. The ontic state can therefore be written as a vector $\vec{c}$ such that the entry $c_i$ in position $i$ of $\vec{c}$, specifies the set of outcomes that we will definitely obtain if we perform the set of measurements belonging to the context labelled by $i$. I will refer to $\vec{c}$ as the \emph{counterfactual outcome}, since it specifies the outcome that we would obtain if we could check  every possible context simultaneously. Of course, since we are dealing with maximal contexts  we can never actually check more than one context, meaning that we can only ever find out one of the entries in the counterfactual outcome.\footnote{ Of course we we can  certainly perform all the measurements associated with one context and then subsequently perform a  measurement from a different context, but by definition the probability distribution over the last measurement will be different from the one that would have been obtained if we had performed it before the other measurements, so we should assume that in general the value of the counterfactual outcome may have changed before we get to the second context} \footnote{In general $\vec{c}$ will be of infinite length, since the set of possible contexts is continuous; however, it may be possible to parametrize $\vec{c}$ using a finite number of parameters - for example, the Kochen-Specker model gives a deterministic model for the pure states of a qubit using only two parameters.}

According to the usual specification of an ontological model, a preparation procedure $P$ leads to a probability distribution $\mu_P$ over the space of counterfactual outcomes $\vec{c}$. I will say that $\mu_P$ is \emph{unbiased} if it is the case that for each $i$, the marginal probability distribution induced by $\mu_P$ over the possible values of $c_i$ is independent of the values of the other entries in the vector $\vec{c}$. Because we can never check more than one context, we will never be able to determine directly whether a distribution is biased or unbiased. However, in some cases we can make inferences. Consider a set of three maximal contexts, each consisting of a single qubit measurement\footnote{Recall that in the single-qubit case no context can contain more than one measurement, so these single measurements are also maximal contexts}:

\begin{multline}M_1 =  \{ | 0 \rangle \langle 0| ,  | 1 \rangle \langle 1 | \} 
\\ M_2 = \{ ( \frac{1}{2} | 0 \rangle + \frac{\sqrt{3}}{2} | 1 \rangle )( \frac{1}{2} \langle 0 | + \frac{\sqrt{3}}{2} \langle 1 | ),   \: \mathbb{I} -  ( \frac{1}{2} | 0 \rangle + \frac{\sqrt{3}}{2} | 1 \rangle )( \frac{1}{2} \langle 0 | + \frac{\sqrt{3}}{2} \langle 1 | ) \}
 \\ M_3 = \{   ( \frac{1}{2} | 0 \rangle - \frac{\sqrt{3}}{2} | 1 \rangle )( \frac{1}{2} \langle 0 | - \frac{\sqrt{3}}{2} \langle 1 | ),   \: \mathbb{I} -  ( \frac{1}{2} | 0 \rangle - \frac{\sqrt{3}}{2} | 1 \rangle )( \frac{1}{2} \langle 0 | - \frac{\sqrt{3}}{2} \langle 1 | ) \} \\\end{multline}

And consider a set of preparations which prepare the following states:

\[ P_1 \rightarrow | 0 \rangle \\
 P_2 \rightarrow  \frac{1}{2} | 0 \rangle + \frac{\sqrt{3}}{2} | 1 \rangle| \\
  P_3 \rightarrow \frac{1}{2} | 0 \rangle - \frac{\sqrt{3}}{2} | 1 \rangle\]

Let the first three entries of the counterfactual outcome $\vec{c}$ contain the outcomes for the measurements $M_1, M_2, M_3$ in that order, and let the outcomes of the measurements $M_1, M_2, M_3 $ be labelled by $0, 1$ with respect to the order in which they are displayed above. Then $\mu_{P_1}$ must assign probability $0$ to all counterfactual outcomes $\vec{c}$ with $c_1 = 1$;  $\mu_{P_2}$ must assign probability $0$ to all counterfactual outcomes $\vec{c}$ with $c_2 = 1$; and $\mu_{P_3}$ must assign probability $0$ to all counterfactual outcomes $\vec{c}$ with $c_3 = 1$.  

Now consider the preparation in which we choose a number  $x$ uniformly at random from $\{ 1, 2, 3 \}$ and then perform the corresponding preparation $P_x$, thus producing the maximally mixed state. Given that this is the maximally mixed state, the probability of obtaining the result $1$ to measurement $M_1$ is $\frac{1}{2}$, and likewise for measurements $M_2$ and $M_3$. However, if we make the standard assumption that distributions over ontic states compose under convex compositions in the same way as other probability distributions, it is clear that there is no possible way for us to obtain an ontic state where we are certain to get the result $1$ to all three of these measurements; that is to say, the corresponding distribution $\mu_{mix;0}$ must assign probability $0$ to all counterfactual outcomes with $[c1, c2, c3] = [1,1,1]$, i.e. $\mu_{mix;0} (c_1 = 1 | c_2 = 1, c_3 = 1) = 0$. However, $\mu_{mix;0}(c_1 = 1)  =  \frac{1}{2}$, and therefore $\mu_{mix;0}$ is biased. 

Moreover, in general  different ways of producing the same mixed state will lead to different biases. For example, suppose we instead prepare the maximally mixed state by making an equal mixture of the states $| 0 \rangle$ and $| 1 \rangle$. Preparing the state $| 1 \rangle$ individually leads to a non-zero probability of obtaining the result $1$ for each of the measurements $M_1$, $M_2$ and $M_3$, so there seems no reason to think that a preparation of $| 1 \rangle$ can't ever prepare an ontic state  with $[c1, c2, c3] = [1,1,1]$. Thus prima facie we might expect that when the maximally mixed state is produced in this way, the resulting distribution $\mu_{mix;1}$ will not satisfy $\mu_{mix;1}(c_1 = 1 | c_2 = 1, c_3 = 1) = 0$ - that is to say, $\mu_{mix;0}$  and $\mu_{mix;1}$ will exhibit different biases. 

Of course, this is not a proof - after all, in principle it's possible that preparing the state  $| 1 \rangle$ actually never produces an ontic state with with $[c1, c2, c3] = [1,1,1]$, since there is no direct way for us to check. However, we can turn this informal argument into a proof by simply rewriting Spekkens' classic proof of preparation contextuality in terms of counterfactual outcomes; this is done in appendix \ref{appprepfine2}. Obviously this change makes the proof less general and thus in a sense less interesting, but working with counterfactual outcomes makes it more straightforward to see  \emph{why} we can't represent these five different preparations of the maximally mixed states by the same probability distribution over ontic states: it's because  they produce different biases, and there is no possible distribution over counterfactual outcomes which respects all the different biases produced by these different preparations. So the results of these preparations are in fact meaningfully different from the point of view of the \emph{counterfactual} outcomes, since they produce different biases, but since we can't perform the measurements for more than one maximal context we can never check these biases by direct measurements, and thus the results of the preparations are operationally indistinguishable. 

This allows us to  express the fine-tuning problem associated with preparation contextuality in a more concrete way:  the question is, why do there exist sets of preparations in quantum mechanics which give rise to distributions over counterfactual outcomes  which are fine-tuned in such a way that the differences between them show up only  in the biases and not in the observable probabilities for measurements within any single maximal context? A similar analysis could be applied in the case of indeterministic response functions; instead of a counterfactual outcome the ontic state would assign a probability distributions over possible outcomes to every possible context, and a `bias' would be defined in terms of correlations between the probability distributions assigned to  different contexts; then again, for sets of preparations exhibiting preparation contextuality, the differences between the results of these preparations would show up only in the biases and not in the observable probabilities for any individual context. Metaphorically speaking this effect looks almost `conspiratorial,' as if the extra information in the counterfactual outcome is being deliberately hidden from us by the universe. Thus, again, there is a strong case to be made that this fine-tuning effect demands some sort of explanation.

Moreover, this analysis demonstrates that the existence of preparation contextuality is closely tied to the existence of distinct maximal contexts, since if we didn't have distinct maximal contexts then we would always be able to access the entire counterfactual outcome for any preparation, meaning that different distributions over counterfactual outcomes would always be operationally distinguishable. This makes it clear that the connection between preparation and measurement contextuality goes beyond OEP - both can exist only because quantum mechanics allows distinct maximal contexts, and both can be understood as fine-tuning problems.

\section{How can fine-tuning be explained? \label{fine}}  

  It has frequently been observed that Kochen-Specker and measurement contextuality are closely related to non-locality - for example, they can be unified mathematically by the sheaf-theoretic framework, which we will discuss in section \ref{logic}. Prima facie this might seem puzzling, because non-locality seems to entail some sort of violation of classical ideas about causality,  whereas contextuality does not seem to require any exotic physical mechanisms: thus, for example, Rudolph and Harrigan opine that contextuality is `an entirely natural requirement of realistic theories, in no way comparable  to the un-intuitive nature of non-locality.'\cite{harrigan2007ontological} So one might be tempted to think that the relationship between contextuality and non-locality is a purely formal one.

   However,  once we start seeing contextuality as a fine-tuning problem, the analogy becomes much clearer. It is now well-recognised that non-local quantum correlations are fine-tuned in the causal modelling sense\cite{SpekkensWood}:  once one accepts the possibility of non-local influences, the puzzling thing is not the dependence of the measurement outcome on the other measurement choices, but rather the fact that this dependence is precisely hidden by the non-signalling property. And similarly, we have just seen that in the case of measurement contextuality the problem is not the dependence on the context per se, but rather the fact that this dependence is precisely hidden by the Gleason property. So fine-tuning is quite a generic feature of quantum correlations, and therefore it should be a priority for any realist account of the quantum world to offer an explanation for this behaviour. The precise form of the fine-tuning will depend on one's account of the nature of the probability distributions over ontic states described by an ontological model: if one takes them to be objective chances, then the relevant distributions will have to be carefully adjusted and appropriately coordinated across spacetime, whereas if one takes them to be subjective probabilities over real distributions which are fixed by past evolution, then presumably one will have to encode the relevant fine-tuned distributions into the initial state of the universe. However in both of these cases the fine-tuning effect seems to demand some sort of explanation. 
  
It should be reinforced that fine-tuning in the causal modelling sense is disanalogous in several important ways from the `fine-tuning' often discussed with reference to naturalness arguments in cosmology and particle physics, such as the  cosmologists' concern that the Cosmological Constant is much smaller than it `ought' to be\cite{RevModPhys.61.1}, and that the particle physicsts' worry that the Higgs mass is not as close to the Standard Model's high energy cutoff as it `ought' to be\cite{pittphilsci11529}. First, both of these arguments are strongly theory-dependent, whereas the fine-tuning arguments around quantum non-locality and contextuality are very deliberately \emph{not} theory-dependent, since they employ a framework which assumes nothing other than the existence of some causal description. Second, concerns about the cosmological constant or the Higgs mass pertain only to the value of a single fixed parameter, so if there \emph{is} fine-tuning involved it affects only one value; whereas in the non-locality and contextuality cases it seems that either the entire initial state of the universe or a collection of probability distributions across the entire course of history must be carefully adjusted to ensure all dependencies exactly average out, so the fine-tuning involved seems much more `conspiratorial.'  I make this point because naturalness and fine-tuning arguments in cosmology and particle physics have come in for a fair amount of criticism in recent years\cite{bianchi2010prejudices,dine2015naturalness,Hossenfelder_2019}, so it is important to be clear that these sorts of criticisms do not necessarily translate to arguments surrounding the fine-tuning exhibited by quantum non-locality and contextuality. 

So what sort of explanation could  be offered for the fine-tuning of these quantum phenomena? Of course one option is to simply accept the existence of fine-tuning, perhaps in the context of some sort of superdeterministic model\cite{10.3389/fphy.2020.00139}, but what if we want to  \emph{eliminate}  the fine-tuning, or at least account naturally for it? 
We might perhaps give a historical `equilibration' account as proposed in some of the literature on causal modelling\cite{dash2005restructuring}. For example, in some versions of the de Broglie-Bohm theory it is suggested that if the de Broglie-Bohm particles start out in a randomly selected initial state, due to properties of the time evolution equations it is likely that they will over time evolve closer and closer to the  `quantum equilibrium' distribution which is necessary to reproduce the empirical results of quantum mechanics\cite{Valentini_2005}. So although the quantum equilibrium distribution is in a sense `fine-tuned,' this fine-tuning is brought about by a natural process of equilibration rather than by directly adjusting the ontic state distributions. This approach does have the consequence that events in the early universe may not have obeyed standard quantum statistics: in particular, if we use this sort of account to explain the fine-tuning effects which enforce no-signalling and Gleason's property, we will have to accept that in the early universe, no-signalling and Gleason's property were sometimes violated\cite{valentini2001signallocality}. Such violations of no-signalling may make committed relativists a little queasy, and would also lead to the existence of problematic phenomena like closed causal loops\cite{QMG}. Moreover, equilibration models amount to putting in by hand a strong form of empirical time-asymmetry for which we currently have no direct evidence; Valentini has made suggestions about where we might look for such evidence\cite{Valentini:2004ep}, but if confirmation is not forthcoming after these searches it would seem appropriate to explore other options.

So is it possible to explain the fine-tuning associated with no-signalling and Gleason's property in a way which ensures that these constraints are \emph{universally} satisfied? By definition we can't give a causal model of these phenomena which is not fine-tuned, and since it is common to conflate `causal' and `realist,' one might worry that this effectively rules out all possible realist explanations.  However, although the causal modelling approach is very general, it does make a few assumptions which we might choose to give up in preference to a wholesale abandonment of realism. Standard causal modelling assumes that causal graphs employed are directed and acyclic, and therefore it presupposes the existence of a unique, well-defined causal order: in particular, the use of directed acyclic graphs entails that it can never be the case that both $X$ causes $Y$ and also $Y$ causes $X$ (either directly or indirectly via the mediation of other variables). Now, we have seen from the results of Cavalcanti and Shrapnel and Costa that simply adopting retrocausality will not suffice to get rid of the fine-tuning: as long as we have a well-defined causal order we will still have a fine-tuning problem, even if that causal order is not aligned with the usual arrow of time. However, as argued in ref \cite{Adlamspooky}, it is important to distinguish between two different forms of retrocausality. The `causal mediation' approach to retrocausality preserves the notion of a well-defined causal order, either simply switching the temporal direction of the arrow of causality, or keeping the `forwards' arrow of causality and adding an additional `backwards' arrow of causality, as in the two-state-vector framework\cite{Aharonovtwostate}. By contrast, in the `all-at-once' approach we postulate models in which the entire history is selected in an atemporal fashion, for example by optimizing some quantity over the whole history, and therefore in this picture we would not expect to see any well-defined causal order. And in the all-at-once picture  it can certainly be the case that both $X$ causes $Y$ and also $Y$ causes $X$, since $X$ and $Y$ must be chosen together to satisfy the constraints of the all-at-once, atemporal models. All-at-once models would therefore be expected to induce  relations which can't in general be expressed as a directed acyclic graph, meaning that the fine-tuning problem cannot even be posed. 

Moreover, within an all-at-once model we can straightforwardly explain apparently `fine-tuned' phenomena by adopting a \emph{teleological} account. That is, rather than employing the `fine-tuned' parameters as an input to the model, we impose the desired property (e.g. non-signalling or Gleason's property) directly as a constraint on the model, and then allow the supposedly fine-tuned parameters to be produced as an  \emph{output} of the model, so the fine-tuning is not an inexplicable `conspiratorial' effect but rather a consequence of the fact that as part of the model the operational statistics are constrained to satisfy the desired property. For example, we could produce a contextual model which obeys Gleason's property by simply imposing as a constraint `the probability for a measurement outcome never depends on the context.'

This sort of constraint is an example of what is known as an `epistemic restriction,' and it has been shown that ontological models with epistemic restrictions can reproduce many features of quantum mechanics\cite{Spekkens_2007, spekkens2016quasi}. However, simply imposing Gleason's property as a constraint looks fairly ad hoc and unmotivated: ideally we would like to come up with some more general reason why the model should be subject to such a constraint. For example, in refs \cite{QMG} the fine-tuned nature of non-local quantum correlations was explained by imposing a global determinism constraint in the form of a prohibition on certain sorts of loop compositions, which entails that all non-local correlations must be non-signalling and so the parameters must be `fine-tuned' to prevent signalling. Is there some similar argument to be made in the contextuality case?

\subsection{Signalling from the future}

In fact, there is very good reason why the universe should not allow violations of Gleason's property: because this would give us a way to perform signalling backwards in time. To see this, suppose we have two maximal contexts  $\{ A, B, C\}$ and $\{ A, D, E\}$ such that the probability for obtaining a positive result to the measurement $\{ A,  \: \mathbb{I} - A\}$ is different in the two contexts. From the definition of a context the probability for obtaining the result $A$ to the measurement $\{ A,  \: \mathbb{I} - A\}$ does not depend on the order in which the measurements within the context are performed, so we may suppose that $\{ A, \mathbb{I} -A\}$ is performed first and the decision about whether to proceed with $\{ \{  B,  \: \mathbb{I} - B\}, \{  C,  \: \mathbb{I} - C\} \}$ or $\{ \{  D,  \: \mathbb{I} - D\}, \{  E,  \: \mathbb{I} - E\} \} $ is made later. Then the probability that we obtain the result $A$ to the measurement $\{ A,  \: \mathbb{I} - A\}$ depends on a future decision about which additional measurements to perform, so the result of that measurement is a `signal' from the future. 

Of course, one might wonder what exactly is wrong with signalling from the future.
As a first response, we might appeal to the problem of `bilking' - an argument due to  Black\cite{10.2307/3326929} which suggests that backwards causation is impossible because after observing the putative effect we can always choose to interfere to prevent the putative cause from happening, so the putative cause can't in fact be the actual cause. There has been much discussion over bilking as it applies to deterministic backwards causation, but note that what we are dealing with here is not deterministic but rather \emph{probabilistic} backwards causation, since in a theory that violates Gleason's property  it need not be the case that we are \emph{guaranteed} to obtain a positive result to the measurement  $\{ A,  \: \mathbb{I} - A\}$ in one context and \emph{guaranteed} to obtain a negative result in the other context; to violate Gleason's property it's enough that the positive result is more likely in one context than the other. And the bilking argument looks significantly weaker in the context of probabilistic backwards causation, as  there is no reason to think that probabilistic relationships will stay the same in a bilking scenario. For example, suppose we know that the outcome $A$ for the measurement  $\{ A,  \: \mathbb{I} - A\}$ is more likely in the context  $\{ A, B, C\}$, and suppose we decide that whenever we obtain the result $A$ to the measurement $\{ A,  \: \mathbb{I} - A\}$ we will go on to perform the measurements $\{ \{  D,  \: \mathbb{I} - D\}, \{  E,  \: \mathbb{I} - E\} \} $. Thus it will come to pass that if we perform repeated instances of this experiment, the relative frequency of obtaining the result $A$ in the context $\{ A, B, C\}$ will be zero, whilst the relative frequency of obtaining the result $A$ in the context $\{ A, D, E\}$ will be one, so the effective relative frequencies in this particular experiment will not match the probabilities obtained from measurements without bilking. This may seem odd, but it doesn't lead to any actual contradiction, since it  certainly need not be the case that probabilistic relationships observed where the context is selected independently should no still obtain when the choice of context ceases to be independent of the measurement results. To make the bilking argument stick here we would presumably have to adopt some fairly robust account of the nature of the probabilities involved, so the argument could easily be rejected by anyone who favours a different account of probability.

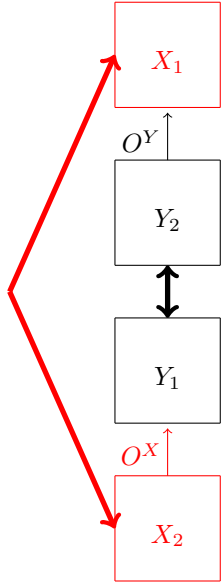
\begin{figure}
	\centering
	\begin{tikzpicture}[scale=0.7]
	
	\coordinate (a) at (0,0);
	\coordinate (b) at (0,2);
	\coordinate (c) at (2,2);
	\coordinate (d) at (2,0);

	\draw[red] (a) -- (b);
	\draw[red] (b) -- (c);
	\draw[red] (c) -- (d);
	\draw[red] (d) -- (a);

	\coordinate (e) at (0,3);
	\coordinate (f) at (0,5);
	\coordinate (g) at (2,5);
	\coordinate (h) at (2,3);

	\draw[black] (e) -- (f);
	\draw[black] (f) -- (g);
	\draw[black] (g) -- (h);
	\draw[black] (h) -- (e);
	
	\coordinate (i) at (0,9);
	\coordinate (j) at (0,11);
	\coordinate (k) at (2,11);
	\coordinate (l) at (2,9);

	\draw[red] (i) -- (j);
	\draw[red] (j) -- (k);
	\draw[red] (k) -- (l);
	\draw[red] (l) -- (i);
	
	\coordinate (m) at (0,6);
	\coordinate (n) at (0,8);
	\coordinate (o) at (2,8);
	\coordinate (p) at (2,6);

	\draw[black] (m) -- (n);
	\draw[black] (n) -- (o);
	\draw[black] (o) -- (p);
	\draw[black] (p) -- (m);

	\draw[red, ->] (1,2) -- (1,2.9);
	\draw[black, ->] (1,8) -- (1,8.9);

	\node[below, red] at (0.5,2.8) {$O^X$};
	\node[below] at (0.5,8.75) {$O^Y$};

	\draw[black, <->, line width=2pt] (1,5) -- (1,6);
	\draw[red, <-, line width=2pt] (0,1) -- (-2,5.5);
	\draw[red, ->, line width=2pt] (-2,5.5) -- (0,10);

	\node[below, red] at (1,1.25) {$X_2$};
	\node[below, red] at (1,10.25) {$X_1$};
	\node[below] at (1,7.25) {$Y_2$};
	\node[below] at (1,4.25) {$Y_1$};
	\end{tikzpicture}	
	\caption{Schematic diagram of the composition of two processes used in the derivation of the Gleason property}
	\label{figsig}
\end{figure}

Alternatively, we can make an argument very similar to that of ref \cite{QMG}, by observing that `backwards signalling' due to violations of Gleason's property could be used to create a closed loop by the arrangement shown in diagram \ref{figsig}. Here the labels $X, Y$ refer to two different physical systems undergoing measurements, $X_1, Y_1$ are black boxes which measure  $\{ \{  B,  \: \mathbb{I} - B\}, \{  C,  \: \mathbb{I} - C\} \} $ if they are given input $0$ and instead measure $\{ \{  D,  \: \mathbb{I} - D\}, \{  E,  \: \mathbb{I} - E\} \} $ if they are given input $1$, and $X_2, Y_2$ are black boxes which implement  $\{ A,  \: \mathbb{I} - A\}$ and then return $0$ if the result is $A$ and $1$ if the result is $  \: \mathbb{I} - A$. As argued in ref \cite{QMG}, the problem with closed loops like this is that the variables inside the loop have no cause outside the loop and therefore the loop is necessarily indeterministic. Thus applying a constraint of global determinism entails that the Gleason property must hold for these measurements, i.e.  if $O$ is the outcome of the measurement $\{ A, \mathbb{I} -A\}$ and $I$ is the choice of     $\{  B,  \: \mathbb{I} - B\}, \{  C,  \: \mathbb{I} - C\}$ or  $\{  D,  \: \mathbb{I} - D\}, \{  E,  \: \mathbb{I} - E\}$, we must have $I(O: I) = 0$. This proof is given in appendix \ref{appfine}. So we can give a teleological explanation for the fine-tuning exhibited in measurement contextuality: in general measurement outcomes may indeed depend on the  context, but a global prohibition on indeterministic loops enforces that these dependencies must average out such that Gleason's property is obeyed.

\subsection{Preparation contextuality \label{prep}} 

Can we give a similar explanation for the case of preparation contextuality? More specifically, can we explain why it should be the case that the distributions over counterfactual outcomes associated with certain sets of convex decompositions are fine-tuned in such a way that the differences between these distributions can't be detected by looking at any individual context?

In fact, this question can be answered by relating the existence of preparation contextuality to the existence of non-local non-signalling correlations. Specifically, the two phenomena are related by the  Choi-Jamio\l{}kowski isomorphism, which describes the mathematical correspondence between quantum channels and entangled bipartite states \cite{Jiang}. Operationally, this means that for any scenario (X) where Alice chooses a measurement from a fixed set of  measurements and performs it on half of an entangled state, and Bob does the same on the other half, there exists a scenario (Y) producing exactly the same operational statistics, in which Alice  selects and performs one preparation $P$ out of some set $\mathbb{C}$ of possible convex decompositions and then subsequently performs a measurement $M$. In ref \cite{AdlamCJ}, it was shown that in any operational theory which exhibits this operationalized version of the Choi-Jamio\l{}kowski isomorphism, the existence of non-local non-signalling correlations entails the existence of preparation contextuality.   
And using this operational version of the isomorphism we can straightforwardly see the reason why we must be unable to detect the differences between the distributions over counterfactual outcomes produced by the various different preparations in $\mathbb{C}$: if it were possible in case (Y) to use the results of $M$ to figure out which preparation from $\mathbb{C}$ was chosen, then in case (X) it would be possible to use the results on one half of the entangled state to infer which measurement was performed on the other half of the entangled state, so we could use this process to perform signalling. Thus in order to preserve no-signalling in case (X), in the isomorphic case (Y) the distributions over counterfactual outcomes must be fine-tuned in order to prevent us from detecting any differences between them. 

Moreoever, in ref \cite{QMG} no-signalling is  derived from a global prohibition on indeterministic closed loops in a construction similar to the one used above. 
So we can indeed give a teleological  explanation for the fine-tuning exhibited in preparation contextuality: in general distributions over ontic states will depend on the particular choice of preparation, but a global prohibition on indeterministic closed loops enforces that for sets of preparations which are mapped by the CJ isomorphism to  measurements on the same entangled state, these dependencies must average out such that the probabilities over subsequent measurement outcomes don't depend on the choice of preparation procedure. 

Furthermore, this approach demonstrates that not only are no-signalling, preparation contextuality and measurement contextuality all essentially the same type of problem, they also all have the same type of solution - they can all be explained as the consequence of a teleological constraint ruling out indeterministic closed loops. This is important, because if we had three similar types of fine-tuning in quantum mechanics which all had to be explained in completely different ways the resulting approach would look fairly patchwork and ad hoc, but in fact the teleological account allows us to give a unifying explanation which derives all three phenomena as consequences of the same global constraint.

\section{Mathematical formalisms for contextuality \label{math}} 

The study of contextuality has led to several powerful mathematical formalisms which allow us to study the structure of contextual theories. In this section, we discuss four such formalisms and consider what they mean for our conceptual understanding of contextuality.

\subsection{Logical Contextuality \label{logic}}

The `logical consistency' approach to contextuality pioneered by Abramsky and Hardy \cite{Abramsky_2012} takes the idea that measurement outcomes are elements of reality one step further: here, measurement outcomes are regarded as boolean variables, so for example, performing a Bell experiment with settings $A$ and $B$ and getting the results $0$ (true) and $1$ (false) is associated with the proposition `$A$ and not $B$.' Thus a probability distribution over measurement outcomes for a Bell scenario can be described as an assignation of probabilities to propositions, meaning that we can derive from classical logic an upper bound on the sum of the probabilities for various sets of measurement outcomes. Non-local quantum correlations violate a logical inequality of this kind, as do Kochen-Specker correlations. 

This is a very interesting mathematical approach which makes it possible to unify contextuality with non-locality and offers a new way of deriving Bell and contextuality inequalities. However, its conceptual significance is less clear. Is it possible that classical logic might be really violated in the quantum world? In one sense, the answer is clearly no - if `possible' is interpreted in the common sense of `permissible according to by classical logic' then violations of classical logic are not possible. But this response may seem somewhat trite, so perhaps a better way to pose the question is to ask whether we could ever have good reason to entertain the hypothesis that classical logic is violated in the quantum world. 

Let us first observe that there are two ways this claim might be interpreted: first, we could say that quantum measurement outcomes are not the sort of thing to which it is appropriate to apply classical logic, or second, we could make the more dramatic claim that classical logic breaks down at the quantum level, as advocated by von Neumann and Birkhoff\cite{10.2307/1968621} and Putnam\cite{Putnam1968-PUTILE} amongst others. To see the difference, suppose that every day I present you with a different selection of breakfast items from which you may select only one, and I treat the propositions `chooses eggs,' `chooses toast,' `chooses Weetabix' and so on as boolean variables. After some number of days I calculate the probabilities for these boolean variables, and I am surprised to find that the probabilities violate a logical inequality. It doesn't seem likely that anyone would suggest this demonstrates the existence of some special non-classical breakfast logic: rather, my description of your choice as a boolean variable wasn't an accurate representation of the physical situation, because in fact your decision about whether to have eggs on any given day may depend what other options I have presented you with. Similarly, in marble-world logical inequalities are violated precisely because it is not correct to think of measurement operators in marble-world as boolean variables, since marble-world measurements do not ask the question `which one of these propositions is true,' but rather `which of these options is preferred.' So the apparent failure of classical logic in the quantum world can be regarded as simply another way of stating the fact that quantum measurement outcomes cannot be understood as expressing the sorts of stable intrinsic facts which are suitable to be described as boolean variables. 

This demonstrates that we don't \emph{have} to respond to the violation of logical inequalities by giving up on classical logic - we can simply question the set-up of the problem as a test of classical logic in the first place. That said, given our conclusions of section \ref{problem}, one might  argue that rejecting classical logic is an appropriate way to explain away the fine-tuning problem:  as we've seen, the natural way to give a non fine-tuned explanation for the fact that quantum measurement outcomes satisfy the Gleason property is to say they express stable intrinsic facts of the sort which can typically be described as boolean variables, so the proponent of non-classical logic might suggest that either we must accept the existence of fine-tuning or we must suppose that some rules of logic are broken in this case. In light of this argument, it might also be remarked that the breakfast and marble-world examples are not really a fair comparison, since these cases don't satisfy Gleason's property and hence don't exhibit the specific behaviour which motivates treating measurement outcomes as boolean variables in the first place.

Now the apparent dilemma is not quite so black and white, as in section \ref{fine} we have argued that there is an alternative way to provide a non-fine-tuned explanation for the Gleason property - i.e. by appeal to teleological constraints in an `all-at-once' picture. Nonetheless the proponent of non-classical logic might contend that their explanation is a better one - for example, someone very strongly attached to the conventional view of time  might prefer to give up classical logic rather than adopt a teleological view. I will not rehearse here all the difficulties that `quantum logic' presents (see ref \cite{maudlin2018labyrinth}), but with respect to this particular argument, my objection to the  `failure of logic' explanation is not that it fails to be the \emph{best} explanation for contextuality, but that it isn't really very explanatory at all: the fact that classical logic doesn't give the right answer in these circumstances is precisely the fact that needs explaining, so simply positing the failure of classical logic isn't adding very much to our understanding. Of course we can't rule out the possibility that  some features of reality which seem in need of explanation actually can't be explained, but presumably we should explore all viable routes before giving up on explanation altogether.

\subsection{Generalisations of Probability \label{general}}

Proofs that quantum mechanics is contextual implicitly depend on the assumption that the probabilities employed by the underlying ontological models obey the  axioms of classical probability. Thus one possible way to avoid these no-go results is to relax one or more of the classical probability axioms, leading to a non-classical probability space in which it is possible to come up with a non-contextual model. A number of such proposals have been made, including extended probability spaces which allow negative probabilities\cite{MUCKENHEIM1986337} or complex probabilities\cite{ivanovic1978complex, youssef1994complex, Srinivasan_1994}, generalized probability spaces\cite{suppes1975logics} which relax the requirement that we should be able to assign probabilities to all conjunctions of events,  upper probability spaces which are subadditive rather than additive on disjoint measurable sets\cite{suppes1991existence}, and quantum measures which are not additive on pairs of events but which are additive on triples of events\cite{sorkin1995quantum}.

As with the case of logical contextuality, there are two possible ways to interpret these proposals. First, we could say that non-classical probabilities are a convenient mathematical shorthand: quantum measurement outcomes are not the kind of thing which can appropriately be described as `elementary events,' and therefore they don't form a classical probability space, but we can come up with a formalism for non-classical probability which allows us to do calculations with quantum measurement outcomes \emph{as if} they are elementary events governed by non-classical probabilities, though this is not to be taken literally. Alternatively, we could say that quantum measurement outcomes really are elementary events and classical probability simply breaks down at the quantum level, so these non-classical probabilities are just as real as classical probabilities. 

If we take the latter approach, there is a further question about whether these new non-classical probabilities are supposed to be subjective or objective. Recall that the concept of probability has two distinct facets: subjective probabilities describe agents' reasonable degrees of belief as governed by Bayesian reasoning, while objective chances are supposed to describe some sort of real, observer-independent property\cite{DASTON1994330}. Now, the difficulty with the subjective probability approach in this instance is that using non-classical probabilities as degrees of belief will lead to behaviour which does not meet our usual standards for reasonableness. For example, it is a well-known result from decision theory that agents who assign negative probabilities will be vulnerable to having a `Dutch Book' made against them, i.e. a set of bets on which they are guaranteed to lose money, so credences of this sort would seem to violate a minimal condition for rationality\cite{sep-dutch-book}. Moreover, Feintzig and Fletcher\cite{Feintzeig_2017} have developed a framework of `weak hidden variable representations' which is intended to subsume all of these proposals for models of quantum mechanics involving non-classical probabilities, and they show that for any attempt to model a contextuality scenario which violates some Kochen-Specker inequalities using a weak hidden variable representation,  this representation will have to violate either Weak Classicality (the requirement that subsets of the event space consisting of mutually orthogonal projection operators spanning the Hilbert space must be classical probability spaces) or No Finite Null Cover (the requirement there is no subset of events which individually have probability zero but whose disjunction is certain to occur). Feintzig and Fletcher argue that violations of WC can't be accepted, since mutually orthogonal operators are all simultaneously measurable and therefore represent classical events which must be governed by classical probabilities; so they conclude that any weak hidden variable representation will have to violate NFNC. They then  show that any agent whose subjective probability assignment violates NFNC is vulnerable to a Dutch Book; so in fact, if we interpret quantum measurement outcomes as elementary events and adopt some sort of non-classical probabilities as our subjective probabilities, \emph{any} choice of non-classical probabilities will lead to beliefs which are paradigmatically irrational. 

So perhaps the non-classical probabilities should instead be regarded as objective probabilities? Well,  `objective chance' as a concept remains philosophically problematic, but it is generally agreed that objective chances (if they exist at all) must play the functional role picked out by Lewis' Principal Principle: objective chances are simply whatever a rational agent would set their credences to\cite{Lewis1980-LEWASG}. That is to say, if $E$ is a proposition stating the objective chance of $A$, and $E$ contains no inadmissible information about the future, then a rational agent who knows $E$ will set their credence for $A$ equal to the objective chance stated in $E$. There are a number of nuances surrounding the issue of  `admissible information,' but these will not concern us here: we need only observe that the Principal Principle requires some degree of mathematical continuity between objective chances and credences, since we are supposed to be able to set them equal to one another. Moreover, credences are subjective probabilities, and we have just argued that in order to be rational, subjective probabilities must obey the axioms of classical probability; so rational agents can set their credences equal to objective chances only if the objective chances also obey the axioms of classical probability. This line of argument suggests that objective chances can be regarded as such only in virtue of  obeying the axioms of classical probability: a mathematical construction which relaxes one of these axioms might be analogous to probability but it cannot \emph{literally} be a form of probability.

\subsubsection{Where do non-classical probabilities come from? \label{nc}} 

These considerations suggest that non-classical probabilities can't straightforwardly be regarded as either subjective or objective probabilities, so let us therefore return to the idea that non-classical probabilities are to be regarded as a convenient mathematical shorthand - that is to say, something about the nature of the reality underlying quantum systems makes it appropriate to describe them using `non-classical probabilities,' although these numbers are not literally probabilities. In this case, appealing to non-classical probabilities is not really an explanation for contextuality: we still need to say what it is about the nature of the reality underlying quantum mechanics which makes the non-classical probabilities useful.   I  will now proceed to suggest such an explanation for the case of negative probabilities. (I do not mean to suggest that this is the only possible explanation; it is merely an example of the sort of account that would be necessary to make sense of the mathematical utility of non-classical probabilities). 

Let us begin from a classical ontological model on $x$ ontic states, so the epistemic state vectors  lie in the space $\mathbb{R}^x$. Suppose that this is a model for a set of quantum operators on a single quantum system - that is to say, an ontological model for a single system where all the probabilities appearing in the epistemic state distributions and response functions obey the axioms of classical probability. Suppose that this ontological model is measurement contextual, and suppose further that, for the reasons discussed in section \ref{fine}, this model is subject to an epistemic constraint: it must satisfy the Gleason property. Then consider some event $E$ which is represented in the ontological model by  two different response functions $\xi_1$ and $\xi_2$ for two different contexts. By stipulation this event must exhibit the same operational statistics in either context, and thus for any possible preparation $P$, the epistemic state vector $\vec{\mu}_P$ resulting from that preparation must satisfy  $\vec{\xi}_1 \cdot \vec{\mu}_P = \vec{\xi}_2 \cdot \vec{\mu}_P$ (here we employ the geometric interpretation of an ontological model - see appendix \ref{ontological}). Thus all allowed epistemic state vectors are orthogonal to $\vec{\xi}_1 - \vec{\xi}_2$, and therefore all allowed  epistemic state vectors lie in a subspace $S$ of the ontic state space of dimension $x - 1$. 

In Appendix \ref{appgeneral}, we show how to use this fact to compress the model into a new quasi-ontological model defined on $x - 1$ quasi-states, where the model is now non-contextual. However, some of the epistemic state vectors and response functions will typically be forced to have negative entries in this new quasi-ontological model, because the intersection of the nonnegative orthant of $\mathbb{R}^x$ (which is a convex cone with $x$ extreme halflines) with the null space of  $\vec{\xi}_1 - \vec{\xi}_2$ (which is a subspace) will always be a cone, but the number of extreme halflines of this cone will be less than $x$ only in special cases, and if the number of extreme halflines is not less than $x $ it will not be possible to choose a basis of  $x - 1$ vectors for the new quasi-space which ensures that the intersection of the nonnegative orthant of $\mathbb{R}^x$ with the subspace lies inside the nonnegative orthant of $\mathbb{R}^{x - 1} $ as defined by the new quasi-states (since this is a convex cone with   $x - 1$ extreme halflines). Thus some of the  epistemic state vectors and response functions which lie in the nonnegative orthant of $\mathbb{R}^x$ will be taken by the compression outside the nonnegative orthant of the new quasi-space, meaning that they will end up having some negative entries. 

This consequence can be avoided  if the original ontological model uses only a limited portion of the nonnegative orthant of $\mathbb{R}^x$, so that only epistemic state vectors and response functions which are taken by the compression to the nonnegative orthant of the new quasi-space   are considered physically real. But due to the continuity of the space of quantum states and operations it's typically the case that ontological models of quantum mechanics use the whole nonnegative orthant and thus generically  we will get negativity under compression. Indeed, the proof of  of ref \cite{Spekkens_2008},  which shows that the presence of KS-contextuality in a theory is equivalent to the presence of negativity in any possible quasi-probability representation of the theory,  allows us to conclude that in the case of quantum mechanics this compression process will  necessarily always take us outside the nonnegative orthant of the final non-contextual space. The argument we have presented here is a useful complement to that of ref \cite{Spekkens_2008} because it demonstrates \emph{why} contextuality and negativity should be connected in this way:   negative quasi-probabilities arise precisely because the existence of contextuality forces the epistemic state vectors to live on a subspace of the ontic state space, leading to a new quasi-ontological model which will not   have all non-negative probability distributions and response functions. 

So we can now appreciate why it might work mathematically to describe quantum measurement outcomes as elementary events governed by non-classical probabilities: given an underlying reality which is entirely classical and obeys the standard classical probability axioms, negative probabilities will arise when we compress the model to a non-contextual subspace. The negative probabilities are therefore not \emph{literally} probabilities, because the events to which they apply are not literally elementary events: these new response functions are mathematical constructions which have significance only as part of the compression process. Moreover, note that in this construction contextuality is not explained by the existence of negative probabilities, but the reverse: it is the existence of contextuality which gives rise to the appearance of negative probabilities when we compress to a non-contextual representation.

 \subsection{Graph Theory \label{graph}}
 
Beginning with ref \cite{Cabello_2014}, it has become popular to represent a contextuality scenario as an `exclusivity graph,' $G$ where every measurement operator is a vertex $v$ of the graph and the edges $E$ of the graph connect  compatible measurements. We define a \emph{probabilistic model} for such a graph  by assigning nonnegative real numbers $p_v \in [0,1]$ to the vertices $v$ in such a way that for any edge $E$ of the graph, $\sum_{v \in E} p_v \leq 1$. Models for a contextuality scenario that obey deterministic contextuality, known as  non-contextual hidden variable (NCHV) models,  are probabilistic models in which  every value $p_v$ is equal to $0$ or $1$. 

Trivially, it is always possible to find a valid NCHV model for any contextuality scenario - for example, we can simply assign the value $0$ to all vertices. However, the situation changes if we stipulate that all the sets of compatible measurements in the graph are maximal, giving what we will describe as a  \emph{maximal} contextuality scenario. The sum of the probabilities  across a maximal set of compatible measurements must be equal to $1$ (since they could all be combined into a single measurement and that measurement would necessarily have some outcome), so a valid probabilistic model for a maximal scenario must satisfy  $\forall E \quad \sum_{v \in E} p_v = 1$, and an  NCHV model must assign the value $1$ to \emph{exactly} one vertex in each edge. It is not always possible to  find even one model fulfilling these constraints: indeed, the Kochen-Specker  theorem may be expressed in this langauge as the statement that at there exist maximal contextuality scenarios which can be realised by quantum-mechanical measurements  for which there exists no valid NCHV model.  

We can quantify the  amount of contextuality exhibited by a model for a given contextuality scenario  by the expectation value of the witness operator $   \Sigma   := \sum_{i = 1}^q p_i$, where $p_i$ is the probability that we will obtain the outcome $i$ if we perform some measurement for which $i$ is a possible outcome, and we sum over all the measurements\cite{Emersonetc}. It can be shown \cite{Cabello_2014} that the expectation value of $\Sigma $ for any valid NCHV model is bounded above by the independence number $\alpha(G)$ of the exclusivity  graph\footnote{The independence number of a weighted graph $G$ is the largest sum of weights assigned to an independent set, i.e. a set such that  no two two vertices in the set are adjacent\cite{graphtheory}.}. On the other hand, in models where the vertices are assigned probabilities derived from a quantum-mechanical representation in terms of density operators and projective measurements, the witness operator is instead upper bounded by   the  Lovasz number $\theta(G)$ of the exclusivity  graph\footnote{ The  Lovasz number of a  graph $G$ with weights $\{ p_v\}$ is defined as $ \sum_v p_v x_v$ where $\{ x_v \}$ is a set of real numbers such that $\sum_v x_v  \frac{(a_1v)^2}{ || a_v ||^2} \leq 1$ for any orthogonal labelling $\{a^v \}$ (where $a_1^v$ denotes the first entry in the vector $a_v$ and $|| a_v ||$ denotes the magnitude of the vector $a_v$). An  assignation of vectors to the vertices of a   graph $G$ is an  orthonormal labelling iff the vectors representing vertices $i$ and $j$ are orthogonal whenever $i$ and $j$ are not adjacent in $G$. }. 
Finally, in models which obey no constraint other than  Exclusivity - the requirement  that the sum of the probabilities assigned to a set of measurement elements such that any two elements in the set are adjacent, and hence simultaneously measurable, is no greater than 1 -  the witness operator is upper bounded by   the  fractional packing number $v_F(G)$ of the graph\footnote{The  fractional packing number of a $q$-vertex graph $G$ with  weights $\{ p_v\}$ is equal to $\max \sum_{i \in \{1, 2 ... q\}} p_i q_i$ where we maximize over all choices of nonnegative numbers $q_i$ subject to the constraint that for any clique $C$ of $G$, $\sum_{i \in C} q_i \leq 1$.
	A clique of a graph is a set of its vertices such that any two distinct vertices in the set are adjacent.}. Now the independence number, Lovasz number and fractional packing number do not in general coincide: in general we have $ \alpha(G) \leq \theta(G) \leq v_F(G)$ and this suggests two interesting questions: first, why is quantum mechanics more contextual than any NCHV theory, and second, why is quantum mechanics not as contextual as  Exclusivity would allow? 

We have already spent some time addressing the first question, but we have not yet addressed the second.  The graph theoretic framework represents each measurement by a unique vertex, and therefore any graph representation of a contextuality scenario necessarily obeys Gleason's property, which means that the teleological arguments given in the previous section aiming to explain Gleason's property do not suffice to explain why quantum theories can't in general achieve the maximal value of the witness operator compatible with Exclusivity. Thus the question raised by the graph-theoretic approach can be made more specific: why are quantum correlations \emph{more} fine-tuned than Gleason's property requires? Answering this question is somewhat outside the scope of this article, so we will merely observe that since we have seen that Gleason's property is something like a temporal analogue of no-signalling, this question is very similar in form to well-known questions about why quantum mechanics is more local than the no-signalling theorem requires\cite{Pawlowski} and why quantum mechanics obeys a stronger monogamy bound than the no-signalling theorem requires\cite{Popescu:2014wva}, and thus it would be interesting to see if the contextuality bound could be explained in a similar way to the Tsirelson bound and/or the quantum monogamy bound (e.g. see refs \cite{Pawlowski, adlam2020tsirelsons}).

 \subsection{Sheaf-Theoretic Contextuality} 
  
  Abramsky and Brandenburger\cite{Abramsky_2011} have developed a powerful mathematical framework unifying contextuality and non-locality in the language of sheaf theory. In this framework, an  empirical model for a contextuality scenario specifies, for every context in the scenario, a joint probability distribution over the outcomes of the measurements included in that context, such that whenever two contexts intersect (i.e. a measurement appears in two contexts), the distributions for the two contexts agree on the intersection (i.e. they assign the same probability distribution over the outcomes of the shared measurement). If it is possible to come up with a single empirical model which specifies a unique probability distribution over the outcomes of each possible measurement in a context-independent way, that model is known as a `global section.'  

In this framework, we say that an empirical model is contextual if it has no global section, i.e. contextuality can be understood as the existence of some obstruction to a global section. In the non-locality case a global section is equivalent to the existence of a `hidden variable' which determines the outcome of measurements on each system independent of choices for measurements on the other system, and therefore measurements which can be used to violate Bell's theorem can't be associated with a global section, so non-locality is a form of contextuality in this framework. 

We can also distinguish several different levels of contextuality in this framework. An empirical model exhibits  probabilistic contextuality if there is no global section such that for all possible contexts, the marginal of the global section on that context is equal to the empirical model on that context. An empirical model exhibits possibilistic contextuality  if there is no global section such that for all possible contexts, the set of measurement outcomes in that context to which the global section assigns non-zero probability is the same as the set of measurement outcomes in that context to which the empirical model assigns non-zero probability. (Possibilistic contextuality was also studied by Simmons et al in ref \cite{simmons2017contextuality}) An empirical model exhibits strong contextuality if there is no way of choosing a subset $S$ of all the measurement outcomes such that for all possible contexts, the support of $S$ on that context is equal to the set of measurement outcomes in that context to which the model assigns non-zero probability. These forms of contextuality define a hierarchy: any possibilistically contextual empirical model is probabilistically contextual, and any strongly contextual empirical model is possibilistically contextual, but the reverse entailments do not hold, so strong contextuality is indeed the `strongest' of these forms of contextuality. Abramsky and Brandenburger demonstrate that probabilistic, possibilistic and strong contextuality can all be exhibited in quantum contextuality scenarios. 

Our discussion in this article has largely focused on probabilistic contextuality, but many of our comments carry over to possibilistic and strong contextuality. We have argued that probabilistic contextuality and non-locality should be understood as a fine-tuning problem: we accept that for a given ontic state the probabilities for various measurement outcomes may depend on context, and then the mandate is to explain why the context-dependence is hidden. The existence of possibilistic and strong contextuality in quantum mechanics demonstrates that in some cases, for a given ontic state not only the probabilities for various measurement outcomes but even the facts about which measurement outcomes are \emph{possible} may depend on context: again, the crucial question is to explain why this particularly strong form of context-dependence is hidden.

Abramsky and Brandenburger use their framework to prove many interesting results - for example, they demonstrate that for any contextuality scenario satisfying no-signalling and Gleason's property, we can always come up with a global section if we are allowed to use negative probabilities. This accords with the discussion of section \ref{general}, where we saw that a contextual ontological model can always be compressed to a non-contextual model on a subspace of the ontic state space if we allow the model to use some negative probabilities.  Abramsky and Brandenburger also point out that the same is true for non-quantum no-signalling correlations like the PR boxes, so the existence of negative probabilities does not suffice to single out quantum mechanics uniquely. Again, this is what we would expect based on section \ref{general}, as we obtained negative probabilities from a general compression process which can be applied to \emph{any} ontological model which exhibits measurement contextuality. 
 
 Abramsky and Brandenburger also suggest that their approach leads to a new way of thinking about the nature of a `context.' They observe that in the study of contextuality it is usual to define contexts as sets of compatible measurements, with incompatibility understood in terms of  `the quantum-mechanical formalism of non-commuting observables,' but they suggest that their work offers a theory-independent approach to incompatibility, where `the incompatibility of certain measurements can be interpreted as the impossibility - in the sense of mathematically provable non-existence - of joint distributions on all measurements which marginalize to yield the observed empirical distributions.' Clearly this way of thinking has something in common with the operational definition of a context that we have used in this paper.   However, the two definitions are not equivalent: this can be seen most clearly by considering the single-qubit case. Here, we have many measurements which are not `compatible' in the sense defined in section \ref{intro}, since pairs of projective measurements do not commute and therefore if we perform two projective measurements in a row on a single qubit the probability distributions over their outcomes will depend on the order. But for a single-qubit system no measurement can appear in more than one context, and therefore there will always exist a global section for measurements on a single qubit, so no single-qubit measurements are `incompatible' in the sense of Abramksy and Brandenburger. This suggests that it may be useful to keep the issue of  `compatibility' separate from facts about the existence of a global section: certainly, the fact that some measurements can't be simultaneously performed is part of what makes it possible to construct contextuality scenarios in quantum mechanics for which there is no global section, but there do exist non-trivial contextuality scenarios which have a global section and therefore we shouldn't conflate the two issues.

 \section{Conclusion}

 We set out in section \ref{problem} to understand the source of the intuition that there is something problematic about contextual models. We argued that the Identity of Indiscernibles is not particularly relevant to this intuition, and that it also can't straightforwardly be regarded as an application of  Einstein's ideas or Ockham's razor. Ultimately, we concluded that the real difficulty with Kochen-Specker and measurement contextuality is the fact that any model which allows a dependence on context whilst preserving Gleason's property will necessarily have to be fine-tuned to hide the context-dependence: thus measurement contextuality is best thought of as a fine-tuning problem. We then argued that preparation contextuality can also be regarded as a fine-tuning problem. 
 
 Because the causal modelling framework used here and in ref \cite{Cavalcanti_2018} to demonstrate that contextuality is linked with fine-tuning is very general and makes no assumptions about the nature of the latent variables involved, it is not straightforward to avoid this problem by simply moving to a different sort of (causal) model. So, assuming we are not willing to abandon realism altogether, we seem to have three options: we could accept the existence of fine-tuning, we could give some sort of historical account which entails that no-signalling and Gleason's property do not hold universally, or we could adopt an account which is realist but non-causal. Thus for any scientific realists who consider that fine-tuning generally demands an explanation and who believe that key features of quantum mechanics should hold universally, there is a strong motivation  to move to a non-causal, \emph{teleological} picture.  We therefore suggest that the contextual nature of quantum theory can be regarded as further evidence for the `all-at-once' picture of lawhood\cite{QMG, adlam2020tsirelsons, Wharton} in which global constraints and teleological explanations are most natural. Here we offered a possible teleological explanation for Gleason's property in terms of a prohibition on  indeterministic closed loops, and we argued that this allows us to provide a unifying explanation of the fine-tuning effects involved with measurement contextuality, preparation contextuality, and non-locality. 
 
 Finally, we discussed a number of mathematical frameworks which have recently been applied to the study of contextuality. We concluded that although non-classical logic and non-classical probabilities are a powerful tool for the study of contextuality, they need not be taken literally. As an example, we demonstrated how non-classical negative probabilities  arise naturally within an  ontological model obeying classical logic and classical probability theory  when the model is compressed to remove contextuality. We also showed that this compression process agrees with conclusions obtained within the sheaf theoretic hierarchy set by Abramsky and Brandenburger, and we noted that the sheaf theoretic framework demonstrates the existence of an even stronger form of fine-tuning associated with contextuality. 
 
 \section{Acknowledgements} 
 
 The mathematical construction used in section \ref{nc}  was partly inspired by discussions with Rob Spekkens. 
 
 \vspace{2mm} 
 
 This publication was made possible through the support of the ID\# 61466 grant from the John Templeton Foundation, as part of the “The Quantum Information Structure of Spacetime (QISS)” Project (qiss.fr). The opinions expressed in this publication are those of the author(s) and do not necessarily reflect the views of the John Templeton Foundation.

 \appendix
 
 \section{Ontological Models \label{ontological}} 
 
 Several of the proofs presented here make use of the framework of ontological models, which we now briefly recapitulate for ease of reference. 
 
In the ontological models framework, each possible preparations is associated with an epistemic state distribution $\mu_P$ over ontic states such that $\mu_P(\lambda)$ gives the probability that the system will end up in the ontic state $\lambda$ when we perform procedure $P$, and each measurement $M$ and outcome  $O$ is associated with a response function $\xi^{k, M}$ such that $\xi^{k, M}(\lambda)$ gives the probability that outcome $k$ will occur when we perform measurement $M$ on a system which is in ontic state $\lambda$. (Ontological models may also model transformations, but we will not be concerned with them here). A valid ontological model must satisfy the following conditions: 
 
 \begin{enumerate} 
 	
 	\item All entries in the epistemic state distributions $\mu_P$ are nonnegative (since they are probabilities)
 	\item The entries in each epistemic state distribution $\mu_P$ sum to $1$ (since each preparation must prepare some ontic state)
 	
 \item  All entries in the response functions $\xi^{k, M}$ are nonnegative (since they are probabilities)
 
 \item For any measurement/context $M$, and any ontic state $\lambda$, we have that $\sum_{k \in M} \xi^{k, M} (\lambda) = 1$ (since every measurement must have some outcome)
 
 \end{enumerate} 

In some of these proofs we will use a geometric interpretation of ontological models,    where epistemic state distributions $\mu_P$ become epistemic state vectors $\vec{\mu}_P$ with each entry in the vector giving the probability assigned to the corresponding ontic state, and response functions $\xi^{k, M}$ become response function vectors $\vec{\xi}^{k, M}$ with each entry in the vector likewise giving the probability for the corresponding ontic state, with all vectors defined on $\mathbb{R}^N$ where  $N$ is the total number of ontic states. In this language, the probability to obtain the outcome $k$ when we perform measurement $M$ on a system which has been prepared in state $P$ is given by $\vec{\mu}_P \cdot \vec{\xi}^{k,M}$, and the conditions above can be rewritten as follows: 

\begin{enumerate} 
	
	\item All entries in the epistemic state vectors  $\vec{\mu}_P$ are nonnegative 
	\item The entries in each epistemic state vector $\vec{\mu}_P$ sum to $1$ 
	
	\item  All entries in the response function vectors $\vec{\xi}^{k, M}$ are nonnegative 
	
	\item For any measurement/context $M$, we have that $\sum_{k \in M} \vec{\xi}^{k, M}  = \vec{1}$, where $\vec{1}$ is the vector in $\mathbb{R}^N$ with all entries equal to one.
	
\end{enumerate} 
 
 \section{Proof for section \ref{prepfine}: fine-tuning \label{appprepfine}}

 \begin{proof}

 As in ref \cite{Cavalcanti_2018}, given an  operational procedure which takes inputs $X$, $Y$ and produces outputs $A$, $B$, we say that this is  a \emph{no-disturbance phenomenon} iff $P(A|XY ) = P(A|X)$ and
 	$P(B|XY ) = P(B|Y )$ for all values of the variables $A,
 	B, X, Y$ for which those conditionals are defined.
 	\vspace{2mm}

 	Let  $\mathbb{S}$ be a prepare-measure scenario: that is, in scenario   $\mathbb{S}$ we select and perform one preparation $P$ out of some set $\mathbb{C}$ of possible preparations, then  perform a measurement $M$ which results in outcome $O$. Suppose that $\mathbb{C}$ is chosen such that no subsequent measurement $M$ can distinguish which preparation from $\mathbb{C}$ has been used. Since no measurement can distinguish which preparation from $\mathbb{C}$ has been used in this scenario,  $O$ is conditionally independent of $P$, i.e. we have  that for any $O, M$, $P(O | M P) = P(O | M)$. We can define a trivial event $\mathbb{I}$ which always occurs, such that $P( \mathbb{I} | M P) = P( \mathbb{I}| P) = 1$. Thus  $\mathbb{S}$ is a no-disturbance phenomenon. 
 	
 	\vspace{2mm} 
 	
 	Consider some preparation contextual ontological model for $\mathbb{C}$ - i.e. a model in which it is not the case that all of the preparations in $\mathbb{C}$ are represented by the same probability distribution over ontic states. We will turn this model into a causal model $\mathbb{M}$ by treating the ontic states as latent variables.  The measurement probabilities in this causal model will be given by $P( O | M P) = \sum_{\lambda} \mu_P(\lambda) \xi^{O, M}(\lambda))$. Because the preparations in $\mathbb{C}$ are not all represented by the same probability distribution over ontic states, we do not have $\mu_P = \mu_{P'}$ for all $P, P'$ and therefore it is not possible to find a unique probability distribution $P(\lambda)$ such that this model can be expressed as $P( O | M P) = \sum_{\lambda} P(\lambda)  \xi^{O, M}(\lambda)$.

 	\vspace{2mm} 
 	Suppose $\mathbb{M}$ is faithful. In ref \cite{Cavalcanti_2018} it is shown that every faithful causal model for a no-disturbance phenomenon is factorisable, so $\mathbb{M}$ can be written in the form $P( O \mathbb{I} | M P) = \sum_{\lambda} P(\lambda) P(O| M, \lambda) P(\mathbb{I} |P \lambda)$, or equivalently $P( O | M P) = \sum_{\lambda} P(\lambda)\xi^{O, M}(\lambda)$. 
 	
 	\vspace{2mm} 
 	
 	But we have already seen that $\mathbb{M}$  can't be written in this form. Thus we have derived a contradiction; so no causal model derived from a preparation contextual ontological model is faithful, and therefore preparation contextual ontological models are necessarily fine-tuned.

 \end{proof}

\section{Proof for section \ref{prepfine}: counterfactual outcomes \label{appprepfine2}}

\begin{proof} 
	
	Consider the following six states:

	\begin{multline} P_1 : \psi_1  = | 0 \rangle \\ P_2 :  \psi_2 = | 1 \rangle \\ P_3 : \psi_3 = \frac{1}{2} | 0 \rangle + \frac{\sqrt{3}}{2} | 1 \rangle \\ P_4 : \psi_4 = \frac{\sqrt{3}}{2} | 0 \rangle - \frac{1}{2} | 1 \rangle \\ P_5 : \psi_5 = \frac{1}{2} | 0 \rangle - \frac{\sqrt{3}}{2} | 1 \rangle \\ P_6 : \psi_6 = \frac{\sqrt{3}}{2} | 0 \rangle + \frac{1}{2} | 1 \rangle \\ \end{multline}

	And consider the following measurements, with outcomes labelled as follows: 
	
	\begin{multline}M_1 =  \{ 0 = | 0 \rangle \langle 0| , 1 = | 1 \rangle \langle 1 | \} 
	\\ M_2 = \{0 =  ( \frac{1}{2} | 0 \rangle + \frac{\sqrt{3}}{2} | 1 \rangle )( \frac{1}{2} \langle 0 | + \frac{\sqrt{3}}{2} \langle 1 | ), 1=   \: \mathbb{I} -  ( \frac{1}{2} | 0 \rangle + \frac{\sqrt{3}}{2} | 1 \rangle )( \frac{1}{2} \langle 0 | + \frac{\sqrt{3}}{2} \langle 1 | ) \}
	\\ M_3 = \{ 0 =  ( \frac{1}{2} | 0 \rangle - \frac{\sqrt{3}}{2} | 1 \rangle )( \frac{1}{2} \langle 0 | - \frac{\sqrt{3}}{2} \langle 1 | ), 1=   \: \mathbb{I} -  ( \frac{1}{2} | 0 \rangle - \frac{\sqrt{3}}{2} | 1 \rangle )( \frac{1}{2} \langle 0 | - \frac{\sqrt{3}}{2} \langle 1 | ) \} \\\end{multline}

	Then let $P_{i,j} : {i, j} \in \{ 1, 2, 3, 4, 5, 6\}$ be the composite preparation procedure where either preparation $P_i$ or $P_j$ is performed with equal probability. Likewise let $P_{i, j, k}  : \{ i, j, k \} \in \{ 1, 2, 3, 4, 5, 6\}$ be the composite preparation procedure where $P_i$, $P_j$ or $P_k$ is performed with equal probability. 	Each of these preparation procedures gives rise to a mixed state, as described in section \ref{prep}; and in fact, the six states have been selected such that all of the preparations $P_{1,2}$, $P_{3,4}$, $P_{5,6}$, $P_{1,3, 5}$ and $P_{2,4,6}$, give rise to the same mixed state, $ \frac{1}{2} | 0 \rangle \langle 0 | + \frac{1}{2} | 1 \rangle \langle 1 | $.

	Let's suppose that all of these preparations lead to the same probability distribution $\mu$ over counterfactual outcomes. As argued in the main text using the preparation $P_{135}$, $\mu$ must assign probability $0$ to all counterfactual outcomes with $[c1, c2, c3] = [1,1,1]$. Rehearsing the same argument for preparation $P_{246}$ shows that $\mu$ must also assign probability $0$ to all counterfactual outcomes with $[c1, c2, c3 ] = [0,0,0]$. 
	
	Now  suppose that the probability of obtaining  $[c1, c2, c3 ] = [1,1,0]$ when we perform $P_5$ is $x$. Clearly the probability  of obtaining  $[c1, c2, c3 ] = [1,1,0]$ when we perform $P_1$, $P_3$ or $P_6$ is zero. So when we perform $P_{56}$, the probability of obtaining $[c1, c2, c3 ] = [1,1,0]$ is $\frac{1}{2}x$, while when we perform $P_{135}$  the probability of obtaining $[c1, c2, c3 ] = [1,1,0]$ is $\frac{1}{3}x$, so if these distributions are the same we have $\frac{1}{2} x = \frac{1}{3} x$, which implies that $x$ is zero. So $\mu$ must assign probability $0$ to all counterfactual outcomes with $[c1, c2, c3 ] = [1,1,,0]$. We can then use the symmetry of the problem to make a similar argument for every other possible combination of counterfactual outcomes, so it turns out that there is no combination of outcomes for $c_1, c_2, c_3$ to which $\mu$ assigns a non-zero probability. But any preparation must produce some counterfactual outcome; so it can't be the case that all of these preparations produce the same distribution over counterfactual outcomes $\mu$.

\end{proof}

 \section{Proof for section \ref{fine} \label{appfine}}
 
 In this section we use methods developed in refs \cite{QMG, adlam2020tsirelsons} to show that a global determinism constraint entails Gleason's property. More specifically, the idea is that  global determinism prohibits the existence of closed loops where the value of some variable in the loop comes `out of nowhere,' i.e. it has no cause outside the loop. Clearly  a world containing such a loop would fail to be deterministic, since the value of the variable in question would not be determined by anything.  This idea is formalised in refs  \cite{QMG, adlam2020tsirelsons} by using the process framework\cite{Oreshkov} and splitting the variables which determine the outcome of the process up into `local controllables' (local matters of fact which influence the outcome, e.g. preparations and measurement settings) and `ontic variables' (anything else which influences the outcome, including possibly non-local facts). We then require that the outcome of a process is fully determined by the values of the local controllables and the ontic variable for the process - e.g. for a process with one local controllable $I$ and one output $O$, we require that $O$ is a function of $I$ and the ontic variable $Q$. We also assume that $I$ and $Q$ are independent. 
 
 One obvious way to preserve determinism in a world containing closed loops would be to say that the value of the variable in the loop is determined by something else in the universe, like the temperature of the sun at a certain time $T$. To rule this possibility out, we demand that processes still function in the usual way when they are composed into loops - so if the outcome of a certain process does not usually depend on the temperature of the sun at time $T$, then when the process is composed into a loop, the  variables within the loop still don't depend on the tempeature of the sun at time $T$. In the process framework, we formalise this by demanding that when two processes $X, Y$ are composed, the outputs $O^X, O^Y$  of the processes are now a function of any remaining local controllables $I^X, I^Y$ plus the ontic variables $Q^X, Q^Y$ for the individual processes - i.e. we are not allowed to use additional information from other ontic variables to determine the values of the outputs. 
 
 We refer to refs \cite{QMG, adlam2020tsirelsons} for  technical details of the definitions for global determinism, processes, controllables and the behaviour of processes under composition. For the purpose of the current proof, it is enough to observe that the composite process depicted in figure \ref{figsig} has no inputs from outside the loop, so there are no local controllables and thus global determinism together with our assumption behaviour under composition entails that the two outcomes $O^X, O^Y$ must be fully determined by the global variables $Q^x, Q^y$ for the two processes, i.e. $H(O^X O^Y| Q^X Q^Y) = 0$. Thus the proof proceeds as follows:

 \begin{proof} 
 	
 	Consider a pair of boxes such that box $1$ takes an input $I \in \{ 0, 1 \}$ and box $2$ produces an output $O$. According to our global determinism assumption, it must be the case that for any implementation of these boxes, $O$ is a function of $I$ and the global variable for this process, $Q$. 
 	
 		\vspace{2mm}
 In this instance, we will say that box $1$  implements either the pair of measurements $\{  B,  \: \mathbb{I} - B\}, \{  C,  \: \mathbb{I} - C\}$  (for input $I = 0$) or  $\{  D,  \: \mathbb{I} - D\}, \{  E,  \: \mathbb{I} - E\}$  (for input $I = 1$) and box $2$ implements the measurement  $\{ A,  \: \mathbb{I} - A\}$, returning outcome $O = 0$ if the result is $A$ and $O = 1$ if the result is $ \: \mathbb{I} - A$. Note that the two measurements are to be made on the same system, and therefore either $X_1$ must be operated in the future lightcone of $X_2$ or vice versa. 
 	\vspace{2mm}
 	
 	Suppose we create two pairs of these boxes, $\{ X_1, X_2 \}$ and $\{ Y_1, Y_2\}$. Then we can perform a composition as depicted in fig \ref{figsig}. That is, we first perform measurement $\{ A,  \: \mathbb{I} - A\}$ on system $X$. Depending on the outcome we then choose to perform either the pair of measurements   $\{  B,  \: \mathbb{I} - B\}, \{  C,  \: \mathbb{I} - C\}$ or the pair of measurements  $\{  D,  \: \mathbb{I} - D\}, \{  E,  \: \mathbb{I} - E\}$ on system $Y$. Then we perform the measurement $\{ A,  \: \mathbb{I} - A\}$ on system $Y$, and depending on the outcome we choose to perform either the pair of measurements   $\{  B,  \: \mathbb{I} - B\}, \{  C,  \: \mathbb{I} - C\}$ or the pair of measurements  $\{  D,  \: \mathbb{I} - D\}, \{  E,  \: \mathbb{I} - E\}$.
 	
 	\vspace{2mm}

 As noted above, if $Q^X$, $Q^Y$ are the global variables associated with these two processes we must have $H(O^X O^Y| Q^x Q^y) = 0$, and hence $H(O^X O^Y) = I(O^X O^Y:Q^X Q^Y) $.

 	\vspace{2mm} 
 	
 	From the definition of the mutual information, $H(O^X O^Y) = H(O^X) + H(O^Y) - I(O^X : O^Y)$ 
 	
 	In this construction, $I(O^X : O^Y) = I(O^X : I^X) = I(O : I)$, so $H(O^X O^Y)= 2H( O) - I(O : I)$,
 	
 	\vspace{2mm} 
 	
 Since $O^X$ is a function of $I^X$ and $Q^X$  and $O^Y$ is a function of $I^Y$ and $Q^Y$, $I( O^X O^Y :Q^X Q^Y) \leq I(O^X I^X  : Q^X ) + I(O^Y I^Y : Q^Y) = 2 I (O I : Q )$.
 	
 	\vspace{2mm} 
 	
 	Since $O$ is a function of $I$ and $Q$, and $I$ and $Q$ are independent,  we have  $H(O ) =  I(O : I)  + I(O I : Q)$ (see lemma 4.3 of ref \cite{QMG}).
 	
 	Combining these results, we obtain: 
 	
 	\begin{equation}2 H(O) - I(O : I) \leq 2 H(\vec{g}) - 2 I (O: I) \end{equation} 
 	
 	\vspace{3mm} 
 	
 	Hence $I(O : I) \leq 0 $. But the Shannon mutual information is nonnegative, so  $I(O : I) = 0$.  
 	
 \end{proof}

\section{Proof for section \ref{nc} \label{appgeneral}}

 Consider a contextuality scenario $\mathscr{S}$ consisting of a set of measurements $\{ M\}$  and contexts $\{ C \}$ such that some measurements appear in more than one context. Suppose that  it is not possible to come up with an ontological model for $\mathscr{S}$ in which each event is represented by a unique response function, i.e. any ontological model for $\mathscr{S}$ must be measurement contextual.  We show that imposing Gleason's property as a constraint on this scenario entails that in any ontological model for the scenario, the allowed preparations produce epistemic state vectors which lie in a subspace $S$ of the ontic state space, with smaller dimension than the full ontic state space. 
 
 \begin{proof}
 	
Consider any ontological model for $\mathscr{S}$. By assumption the model is measurement contextual, so it's possible to find a measurement $\{ M,  \: \mathbb{I} - M \}$ such that performing this measurement and obtaining the outcome $M$ is represented in the ontological model by  two different response functions $\vec{\xi}^{M, C_1}$ and $\vec{\xi}^{M, C_2}$ in two different contexts $C_1$, $C_2$. 
 	
  For a preparation $P$ which is associated with the epistemic state vector $\mu_P$, the probability of obtaining outcome $M$ when we perform $\{ M,  \: \mathbb{I} - M \}$ is $\vec{\xi}^{M, C_1} \cdot \mu_P $,  and the probability of obtaining outcome $M$ when we perform $M_2$ is $\vec{\xi}^{M, C_2} \cdot \mu_P$. Since $\mathscr{S}$  obeys Gleason's property, for any allowed $\mu_P$ these probabilities must be the same. Thus for any allowed preparation $P$, the associated epistemic state vector $\mu_P$  must satisfy  $\vec{\xi}^{M, C_1}\cdot \mu_P = \vec{\xi}^{M, C_2} \cdot \mu_P$;  	thus all allowed epistemic state vectors are orthogonal to $\vec{\xi}^{M, C_1}- \vec{\xi}^{M, C_2}$, and therefore all allowed  epistemic state vectors lie in a subspace $S$ of the ontic state space. 
  
 Note  that if the vectors $\vec{x}$ and $\vec{y}$ are orthogonal to  $\vec{\xi}_1 - \vec{\xi}_2$, then every linear combination $\alpha \vec{x} + \beta \vec{y}$ is also orthogonal to $\vec{\xi}_1 - \vec{\xi}_2$, so $S$ is indeed a subspace.   
 	
 	\end{proof}
 	
 We can project the response functions of the ontological model into the subspace $S$ without changing any of the dot products between probability distributions and response functions, so the predictions of the model will be identical. We show that this projection maps the two response functions $\vec{\xi}^{M, C_1}- \vec{\xi}^{M, C_2}$ to a unique vector in the space $S$. 
 	
 	\begin{proof} 
 	
 	The projection of $\vec{\xi}^{M, C_1}$ on $S$ is: 
 	
 	\begin{multline}\vec{\xi}^{M, C_1} - \frac{ (\vec{\xi}^{M, C_1} \cdot (\vec{\xi}^{M, C_1} - \vec{\xi}^{M, C_2})) (\vec{\xi}^{M, C_1} - \vec{\xi}^{M, C_2})}{ |\vec{\xi}^{M, C_1} - \vec{\xi}^{M, C_2} |^2}  \\
 	\vec{\xi}^{M, C_1} (1 - \frac{\vec{\xi}^{M, C_1}\cdot \vec{\xi}^{M, C_1} - \vec{\xi}^{M, C_1} \cdot \vec{\xi}^{M, C_2}}{\vec{\xi}^{M, C_1} \cdot \vec{\xi}^{M, C_1} - 2 \vec{\xi}^{M, C_1} \cdot \vec{\xi}^{M, C_2} +  \vec{\xi}^{M, C_2} \cdot \vec{\xi}^{M, C_2}}) + \vec{\xi}^{M, C_2} \frac{ \vec{\xi}^{M, C_1} \cdot (\vec{\xi}^{M, C_1} - \vec{\xi}^{M, C_2})}{ |\vec{\xi}^{M, C_1} - \vec{\xi}^{M, C_2} |^2}  \\
 	= \vec{\xi}^{M, C_1} ( \frac{ \vec{\xi}^{M, C_2} \cdot \vec{\xi}^{M, C_2}   - \vec{\xi}^{M, C_1} \cdot \vec{\xi}^{M, C_2} }{ |\vec{\xi}^{M, C_1} - \vec{\xi}^{M, C_2} |^2} ) + \vec{\xi}^{M, C_2} ( \frac{ \vec{\xi}^{M, C_1} \cdot \vec{\xi}^{M, C_1} - \vec{\xi}^{M, C_1} \cdot \vec{\xi}^{M, C_2}}{|\vec{\xi}^{M, C_1} - \vec{\xi}^{M, C_2} |^2} )
 	\\ \end{multline} 
 	
 	From the symmetry of of this equation, it is clear that the projection of  $\vec{\xi}^{M, C_2}$ on $S$ will give the same result.

 	\end{proof} 
 	
 	Now, if after this projection there is still some other measurement operator which is represented by two different response functions for two different contexts, we can perform the same procedure again. The resulting subset of the ontic state space will still be a subspace, because the intersection of two subspaces is a subspace. So we can continue performing this procedure until eventually we have a subspace $S'$ such that  each measurement operator corresponds to a unique vector in  $S'$, i.e. there is no longer any measurement contextuality.

 We can then obtain an orthogonal basis for the subspace $S'$, and use it to come up with a new quasi-ontological model using the components of the basis as the new ontic states. 
 
 \begin{proof}

 	Let $n$ be the dimension of the new subspace $S'$, where $n$ is strictly smaller than the dimension of the original ontic state space $S$. We define an orthogonal basis $\{ \vec{f}^{(\lambda)} \} $  for the subspace $S'$, the $n$ elements of which are associated with the $n$  ontic states of the new model. We will normalise the elements of the basis such that $ |\vec{f}^{(\lambda)} |_1 = 1$, i.e. the sum of the entries is equal to one.

 	Then we may define the (quasi) distributions and (quasi) response functions of the new ontological model as vectors on the space $\mathbb{R}^n$, as follows:
 	\begin{equation}
 	\vec{\mu}_n^{P} (\lambda) \equiv 
 	\vec{\mu}^P \cdot  \vec{f}^{(\lambda)}
 	\end{equation}

 	and 
 	\begin{equation}
 	\vec{\xi}_n^{M}( \lambda) \equiv 
 	\vec{\xi}^{M, C} \cdot \vec{f}^{(\lambda)} 
 	\end{equation}
 	
 	where $C$ is any context that includes $M$; since the compression has eliminated measurement contextuality, any choice of $M$ will lead to the same vector $	\vec{\xi}^{M}$. 

\vspace{2mm}
 	It is clear that this model will reproduce the predictions of the old model. We now show that it also satisfies some of the necessary conditions for a valid ontological model. First, we show that  the new (quasi) response functions are such that for any context, the sum of the response functions associated with obtaining the result $M$ to each of the measurements of the form $\{ M,  \: \mathbb{I} - M\}$ such that $M \in C$ sum to $\vec{1}$:
 	
 	\begin{proof}
 		
 		For any context $C$, and any index $i$, we have: 
 		
 		\begin{multline} \sum_{M \in C} (\vec{\xi}_n^{M  })_i = \sum_{M \in C} \vec{f}^{(\lambda_i)} \cdot \vec{\xi}^{M, C} \\= \sum_j \sum_{M \in C}  \vec{f}^{(\lambda_i)}_j (\vec{\xi}^{M, C})_j \\ = \sum_j \vec{f}^{(\lambda_i)}_j  \sum_{M \in C}  (\vec{\xi}^{M, C})_j\\ \end{multline} 
 		
 We know that the original response functions must satisfy $ \sum_{M \in C} \vec{\xi}^{M, C} = \vec{1}$, since each context represents a measurement whose possible outcomes are given by $M : M \in C$, and hence: 
 		
 		\begin{equation} \sum_{M \in C} (\vec{\xi}^{M, C})_i = \sum_j \vec{f}^{(\lambda_i)}_j = 1 \end{equation} 
 		
 		Hence for any context $C$  we have $\sum_{M \in  C} (\vec{\xi}_n^{M}) = \vec{1}^{n}$ where  $ \vec{1}^{n}$ is the vector on $\mathbb{R}^n$ with all entries equal to one. 
 		
 	\end{proof} 
 	
 	We also show that the new  epistemic state vectors $\{ 	\vec{\mu}_n^{P} \}$ sum to $1$.
 	
 	\begin{proof}

 		For any  vector $\vec{v}$ in the subspace $S$, we have: 
 		
 		\begin{multline} 
 		\sum_j \vec{v}_j = \sum_{i,j} ( \vec{f}^{(\lambda_i)} \cdot \vec{v} ) \vec{f}^{(\lambda_i)}_j \\ = \sum_{i}( \vec{f}^{(\lambda_i)} \cdot \vec{v} ) |\vec{f}^{(\lambda_i)}|_1  \\ = \vec{v} \cdot \sum_{i} \vec{f}^{(\lambda_i)} \\\end{multline} 
 		
 		Hence for every vector $v$ in $S$, we have $\vec{v} \cdot \sum_{i} \vec{f}^{(\lambda_i)} = \sum_j  \vec{v}_j$ and therefore $\sum_i \vec{f}^{(\lambda_i)}  = vec{1}^n$.
 		
 		Thus for any epistemic state vector $	\vec{\mu}_n^{P} $ we must have: 
 		
 		\begin{multline} \sum_i (	\vec{\mu}_n^{P} )_i = \sum_{i,j} (	\vec{\mu}^{P} )_j f^{(\lambda_i)}_j \\ = \sum_j (	\vec{\mu}^{P} )_j (  \vec{1})_j  \\  = \sum_j (\vec{\mu}^P)_j = 1\\  \end{multline}

 	\end{proof}
 	
 	So the new quasi ontological model satisfies all the conditions for an ontological model except possibly the requirement that the response functions and probability distributions should be nonnegative.  In fact  the new probability distributions and response functions  will generally have negative entries. For at each step in this process we compress some space $\mathbb{R}^x$ to a subspace $\mathbb{R}^{x - 1}$ which is the null space of some vector;  the intersection of the nonnegative orthant of $\mathbb{R}^x$ (which is a convex cone with $x$ extreme halflines) with the null space of  $\vec{\xi}_1 - \vec{\xi}_2$ (which is a subspace) will always be a cone, but the number of extreme halflines of this cone will not in general be less than $x$, and if the number of extreme halflines is not less than $x $ it will not be possible to choose a basis of  $x - 1$ vectors for the new quasi-space which ensures that the intersection of the nonnegative orthant of $\mathbb{R}^x$ with the subspace lies inside the nonnegative orthant of $\mathbb{R}^{x - 1} $ as defined by the new quasi-states (since this is a convex cone with   $x - 1$ extreme halflines). Thus  some of the  epistemic state vectors and response functions which lie in the nonnegative orthant of $\mathbb{R}^x$ will be taken by the compression outside the nonnegative orthant  of $\mathbb{R}^{x - 1} $, meaning that they will end up having some negative entries.

	To avoid the appearance of negativity in this process, we would have to insist that only a limited subset of the mathematically possible epistemic state vectors and response functions lying in the nonnegative orthant of the space $\mathbb{R}^x$ correspond to real physically possible states and measurements - specifically, we would have to restrict ourselves to epistemic state vectors and response functions which are taken to the nonnegative orthant of the new quasi-space under compression. But due to the continuity of the space of quantum states and operations it's typically the case that ontological models of quantum mechanics use the whole nonnegative orthant and thus when compressing an ontological model of \emph{quantum} systems we will in general end up with negative entries in some epistemic state vectors and/or response functions. 
	
Finally, note that as a result of the compression process, each event is represented by a unique vector in $S'$, so the new quasi-ontological model is measurement non-contextual. Thus the compression process produces a non-contextual, discrete quasi-probability representation of the theory on a number of quasi-states smaller than the original number of ontic states.  
 	
 \end{proof}

 \bibliographystyle{unsrt}
 \bibliography{newlibrary11}{} 

\begin{thebibliography}{10}

\bibitem{Cabello_2014}
Adán Cabello, Simone Severini, and Andreas Winter.
\newblock Graph-theoretic approach to quantum correlations.
\newblock {\em Physical Review Letters}, 112(4), Jan 2014.

\bibitem{MUCKENHEIM1986337}
W~Mückenheim, G~Ludwig, C~Dewdney, P.R Holland, A~Kyprianidis, J.P Vigier,
  N~{Cufaro Petroni}, M.S Bartlett, and E.T Jaynes.
\newblock A review of extended probabilities.
\newblock {\em Physics Reports}, 133(6):337--401, 1986.

\bibitem{Feintzeig_2017}
Benjamin~H. Feintzeig and Samuel~C. Fletcher.
\newblock On noncontextual, non-kolmogorovian hidden variable theories.
\newblock {\em Foundations of Physics}, 47(2):294–315, Jan 2017.

\bibitem{Abramsky_2011}
Samson Abramsky and Adam Brandenburger.
\newblock The sheaf-theoretic structure of non-locality and contextuality.
\newblock {\em New Journal of Physics}, 13(11):113036, Nov 2011.

\bibitem{spekkens2019ontological}
Robert~W. Spekkens.
\newblock The ontological identity of empirical indiscernibles: Leibniz's
  methodological principle and its significance in the work of einstein, 2019.

\bibitem{Acuna2021-ACUMHV}
Pablo~Acu\ na.
\newblock Must hidden variables theories be contextual? kochen \& specker meet
  von neumann and gleason.
\newblock {\em European Journal for Philosophy of Science}, 11(2):1--30, 2021.

\bibitem{rudolph2006ontological}
Terry Rudolph.
\newblock Ontological models for quantum mechanics and the kochen-specker
  theorem.
\newblock {\em arXiv preprint quant-ph/0608120}, 2006.

\bibitem{harrigan2007ontological}
Nicholas Harrigan and Terry Rudolph.
\newblock Ontological models and the interpretation of contextuality, 2007.

\bibitem{QMG}
Emily {Adlam}.
\newblock {Quantum Mechanics and Global Determinism}.
\newblock {\em Quanta}, 7:40--53, 2018.

\bibitem{adlam2020tsirelsons}
Emily Adlam.
\newblock Tsirelson's bound and the quantum monogamy bound from global
  determinism, 2020.

\bibitem{Wharton}
K.~{Wharton}.
\newblock {The Universe is not a Computer}.
\newblock In {Foster}~B. {Aguirre}, A. and G.~{Merali}, editors, {\em
  Questioning the Foundations of Physics}, pages 177--190. Springer, November
  2015.

\bibitem{Cavalcanti_2018}
Eric~G. Cavalcanti.
\newblock Classical causal models for bell and kochen-specker inequality
  violations require fine-tuning.
\newblock {\em Physical Review X}, 8(2), Apr 2018.

\bibitem{10.2307/24900629}
ANDREW~M. GLEASON.
\newblock Measures on the closed subspaces of a hilbert space.
\newblock {\em Journal of Mathematics and Mechanics}, 6(6):885--893, 1957.

\bibitem{KochenSpecker}
Simon Kochen and E.P. Specker.
\newblock The problem of hidden variables in quantum mechanics.
\newblock In C.A. Hooker, editor, {\em The Logico-Algebraic Approach to Quantum
  Mechanics}, volume~5a of {\em The University of Western Ontario Series in
  Philosophy of Science}, pages 293--328. Springer Netherlands, 1975.

\bibitem{von2018mathematical}
J.~von Neumann, R.T. Beyer, and N.A. Wheeler.
\newblock {\em Mathematical Foundations of Quantum Mechanics: New Edition}.
\newblock Princeton University Press, 2018.

\bibitem{Bub_2010}
Jeffrey Bub.
\newblock Von neumann’s “no hidden variables” proof: A re-appraisal.
\newblock {\em Foundations of Physics}, 40(9-10):1333–1340, Jun 2010.

\bibitem{Jammer1974-JAMTPO-10}
Max Jammer.
\newblock {\em The Philosophy of Quantum Mechanics}.
\newblock New York: Wiley, 1974.

\bibitem{dieks2018von}
Dennis Dieks.
\newblock Von neumann's impossibility proof: Mathematics in the service of
  rhetorics, 2018.

\bibitem{Spekkensepistemic}
R.~W. {Spekkens}.
\newblock {Evidence for the epistemic view of quantum states: A toy theory}.
\newblock {\em Phys Rev A}, 75(3):032110, March 2007.

\bibitem{Spekkenscontextuality}
R.~W. {Spekkens}.
\newblock {Contextuality for preparations, transformations, and unsharp
  measurements}.
\newblock {\em Physical Review A}, 71(5):052108, May 2005.

\bibitem{schmid2020unscrambling}
David Schmid, John~H. Selby, and Robert~W. Spekkens.
\newblock Unscrambling the omelette of causation and inference: The framework
  of causal-inferential theories, 2020.

\bibitem{sep-identity-indiscernible}
Peter Forrest.
\newblock {The Identity of Indiscernibles}.
\newblock In Edward~N. Zalta, editor, {\em The {Stanford} Encyclopedia of
  Philosophy}. Metaphysics Research Lab, Stanford University, winter 2020
  edition, 2020.

\bibitem{Einstein}
A.~Einstein.
\newblock Quantum mechanics and reality.

\bibitem{Valentini_2005}
Antony Valentini and Hans Westman.
\newblock Dynamical origin of quantum probabilities.
\newblock {\em Proceedings of the Royal Society A: Mathematical, Physical and
  Engineering Sciences}, 461(2053):253–272, Jan 2005.

\bibitem{valentini2019foundations}
Antony Valentini.
\newblock Foundations of statistical mechanics and the status of the born rule
  in de broglie-bohm pilot-wave theory, 2019.

\bibitem{Durr_1992}
Detlef Durr, Sheldon Goldstein, and Nino Zanghi.
\newblock Quantum equilibrium and the origin of absolute uncertainty.
\newblock {\em Journal of Statistical Physics}, 67(5-6):843–907, Jun 1992.

\bibitem{Carroll2}
S.~M. {Carroll}.
\newblock {In What Sense Is the Early Universe Fine-Tuned?}
\newblock {\em ArXiv e-prints}, June 2014.

\bibitem{2015arXiv151003706A}
D.~{Almada}, K.~{Ch'ng}, S.~{Kintner}, B.~{Morrison}, and K.~B. {Wharton}.
\newblock {Are Retrocausal Accounts of Entanglement Unnaturally Fine-Tuned?}
\newblock {\em ArXiv e-prints}, October 2015.

\bibitem{sep-fine-tuning}
Simon Friederich.
\newblock {Fine-Tuning}.
\newblock In Edward~N. Zalta, editor, {\em The {Stanford} Encyclopedia of
  Philosophy}. Metaphysics Research Lab, Stanford University, winter 2018
  edition, 2018.

\bibitem{holland1995quantum}
P.R. Holland.
\newblock {\em The Quantum Theory of Motion: An Account of the de Broglie-Bohm
  Causal Interpretation of Quantum Mechanics}.
\newblock Cambridge University Press, 1995.

\bibitem{tastevin2021outcomes}
G.~Tastevin and F.~Laloë.
\newblock The outcomes of measurements in the de broglie-bohm theory, 2021.

\bibitem{Spekkens}
R.~W. {Spekkens}.
\newblock {Contextuality for preparations, transformations, and unsharp
  measurements}.
\newblock {\em Phys Rev A}, 71(5):052108, May 2005.

\bibitem{Shrapnel_2018}
Sally Shrapnel and Fabio Costa.
\newblock Causation does not explain contextuality.
\newblock {\em Quantum}, 2:63, May 2018.

\bibitem{SpekkensWood}
C.~J. {Wood} and R.~W. {Spekkens}.
\newblock {The lesson of causal discovery algorithms for quantum correlations:
  causal explanations of Bell-inequality violations require fine-tuning}.
\newblock {\em New Journal of Physics}, 17(3):033002, March 2015.

\bibitem{RevModPhys.61.1}
Steven Weinberg.
\newblock The cosmological constant problem.
\newblock {\em Rev. Mod. Phys.}, 61:1--23, Jan 1989.

\bibitem{pittphilsci11529}
Porter Williams.
\newblock Naturalness, the autonomy of scales, and the 125 gev higgs, June
  2015.

\bibitem{bianchi2010prejudices}
Eugenio Bianchi and Carlo Rovelli.
\newblock Why all these prejudices against a constant?, 2010.

\bibitem{dine2015naturalness}
Michael Dine.
\newblock Naturalness under stress, 2015.

\bibitem{Hossenfelder_2019}
Sabine Hossenfelder.
\newblock Screams for explanation: finetuning and naturalness in the
  foundations of physics.
\newblock {\em Synthese}, Sep 2019.

\bibitem{10.3389/fphy.2020.00139}
Sabine Hossenfelder and Tim Palmer.
\newblock Rethinking superdeterminism.
\newblock {\em Frontiers in Physics}, 8:139, 2020.

\bibitem{dash2005restructuring}
Denver Dash.
\newblock Restructuring dynamic causal systems in equilibrium.
\newblock In {\em AISTATS}. Citeseer, 2005.

\bibitem{valentini2001signallocality}
Antony Valentini.
\newblock Signal-locality and subquantum information in deterministic
  hidden-variables theories, 2001.

\bibitem{Valentini:2004ep}
Antony Valentini.
\newblock {Black holes, information loss, and hidden variables}.
\newblock 7 2004.

\bibitem{Adlamspooky}
Emily Adlam.
\newblock {Spooky Action at a Temporal Distance}.
\newblock {\em Entropy}, 20(1):41, 2018.

\bibitem{Aharonovtwostate}
Y.~{Aharonov}, E.~{Cohen}, E.~{Gruss}, and T.~{Landsberger}.
\newblock {Measurement and Collapse within the Two-State-Vector Formalism}.
\newblock {\em ArXiv e-prints}, June 2014.

\bibitem{Spekkens_2007}
Robert~W. Spekkens.
\newblock Evidence for the epistemic view of quantum states: A toy theory.
\newblock {\em Physical Review A}, 75(3), Mar 2007.

\bibitem{spekkens2016quasi}
Robert~W Spekkens.
\newblock Quasi-quantization: classical statistical theories with an epistemic
  restriction.
\newblock In {\em Quantum Theory: Informational Foundations and Foils}, pages
  83--135. Springer, 2016.

\bibitem{10.2307/3326929}
Max Black.
\newblock Why cannot an effect precede its cause?
\newblock {\em Analysis}, 16(3):49--58, 1956.

\bibitem{Jiang}
Min Jiang, Shunlong Luo, and Shuangshuang Fu.
\newblock Channel-state duality.
\newblock {\em Phys. Rev. A}, 87:022310, Feb 2013.

\bibitem{AdlamCJ}
E.~Adlam.
\newblock The operational choi–jamiołkowski isomorphism.
\newblock {\em Entropy}, 22(1063), 2020.

\bibitem{Abramsky_2012}
Samson Abramsky and Lucien Hardy.
\newblock Logical bell inequalities.
\newblock {\em Physical Review A}, 85(6), Jun 2012.

\bibitem{10.2307/1968621}
Garrett Birkhoff and John~Von Neumann.
\newblock The logic of quantum mechanics.
\newblock {\em Annals of Mathematics}, 37(4):823--843, 1936.

\bibitem{Putnam1968-PUTILE}
Hilary Putnam.
\newblock Is logic empirical?
\newblock {\em Boston Studies in the Philosophy of Science}, 5, 1968.

\bibitem{maudlin2018labyrinth}
Tim Maudlin.
\newblock The labyrinth of quantum logic, 2018.

\bibitem{ivanovic1978complex}
ID~Ivanovi{\'c}.
\newblock On complex bell’s inequality.
\newblock {\em Lettere al Nuovo Cimento (1971-1985)}, 22(1):14--16, 1978.

\bibitem{youssef1994complex}
Saul Youssef.
\newblock Is complex probability theory consistent with bell's theorem?
\newblock {\em arXiv preprint hep-th/9406184}, 1994.

\bibitem{Srinivasan_1994}
S~K Srinivasan and E~C~G Sudarshan.
\newblock Complex measures and amplitudes, generalized stochastic processes and
  their applications to quantum mechanics.
\newblock {\em Journal of Physics A: Mathematical and General}, 27(2):517--537,
  jan 1994.

\bibitem{suppes1975logics}
Patrick Suppes.
\newblock Logics appropriate to empirical theories.
\newblock In {\em The Logico-Algebraic Approach to Quantum Mechanics}, pages
  329--340. Springer, 1975.

\bibitem{suppes1991existence}
Patrick Suppes and Mario Zanotti.
\newblock Existence of hidden variables having only upper probabilities.
\newblock {\em Foundations of Physics}, 21(12):1479--1499, 1991.

\bibitem{sorkin1995quantum}
Rafael~D Sorkin.
\newblock Quantum measure theory and its interpretation.
\newblock {\em arXiv preprint gr-qc/9507057}, 1995.

\bibitem{DASTON1994330}
Lorraine Daston.
\newblock How probabilities came to be objective and subjective.
\newblock {\em Historia Mathematica}, 21(3):330 -- 344, 1994.

\bibitem{sep-dutch-book}
Susan Vineberg.
\newblock {Dutch Book Arguments}.
\newblock In Edward~N. Zalta, editor, {\em The {Stanford} Encyclopedia of
  Philosophy}. Metaphysics Research Lab, Stanford University, spring 2016
  edition, 2016.

\bibitem{Lewis1980-LEWASG}
David Lewis.
\newblock A subjectivist's guide to objective chance.
\newblock In Richard~C. Jeffrey, editor, {\em Studies in Inductive Logic and
  Probability}, pages 83--132. University of California Press, 1980.

\bibitem{Spekkens_2008}
Robert~W. Spekkens.
\newblock Negativity and contextuality are equivalent notions of
  nonclassicality.
\newblock {\em Physical Review Letters}, 101(2), Jul 2008.

\bibitem{Emersonetc}
J.~{Emerson}, D.~{Serbin}, C.~{Sutherland}, and V.~{Veitch}.
\newblock {The whole is greater than the sum of the parts: on the possibility
  of purely statistical interpretations of quantum theory}.
\newblock {\em ArXiv e-prints}, December 2013.

\bibitem{graphtheory}
J.~{Bondy} and U.~{Murty}.
\newblock {\em {Graph Theory With Applications}}.
\newblock Elsevier Science Publishing, 1976.

\bibitem{Pawlowski}
M.~{Pawlowski}, T.~{Paterek}, D.~{Kaszlikowski}, V.~{Scarani}, A.~{Winter}, and
  M.~{{\.Z}ukowski}.
\newblock {Information causality as a physical principle}.
\newblock {\em Nature}, 461:1101--1104, October 2009.

\bibitem{Popescu:2014wva}
Sandu Popescu.
\newblock {Nonlocality beyond quantum mechanics}.
\newblock {\em Nature Phys.}, 10(4):264--270, 2014.

\bibitem{simmons2017contextuality}
Andrew~W Simmons, Joel~J Wallman, Hakop Pashayan, Stephen~D Bartlett, and Terry
  Rudolph.
\newblock Contextuality under weak assumptions.
\newblock {\em New Journal of Physics}, 19(3):033030, 2017.

\bibitem{Oreshkov}
O.~{Oreshkov} and C.~{Giarmatzi}.
\newblock {Causal and causally separable processes}.
\newblock {\em New J. Phys}.

\end{thebibliography}
 
 \end{document}